\documentclass[iop,natbib]{emulateapj}
\usepackage{amsmath,amssymb}
\usepackage{epstopdf}

\newcommand{\mytitle}{{ELECTRON PRE-ACCELERATION AT NONRELATIVISTIC HIGH-MACH-NUMBER PERPENDICULAR SHOCKS}}

\newcommand{\jn}{}
\newcommand{\mpo}{ }
\newcommand{\mpon}{}
\newcommand{\ab}{}
\newcommand{\jnn}{}

\newcommand{\mponn}{}
\newcommand{\jnr}{}
\newcommand{\abr}{}

\usepackage{natbib}
\usepackage{comment}
\usepackage{color}

\shorttitle{Non-relativistic perpendicular shocks}
\shortauthors{Bohdan, Niemiec, Kobzar, \& Pohl}
\begin{document}

\title{\mytitle}

\author{Artem Bohdan\altaffilmark{1}, Jacek Niemiec\altaffilmark{1}, Oleh Kobzar\altaffilmark{1}, Martin Pohl\altaffilmark{2,3}}
\email{artem.bohdan@ifj.edu.pl}
\altaffiltext{1}{Instytut Fizyki J\c{a}drowej PAN, ul. Radzikowskiego 152, 31-342 Krak\'{o}w, Poland}
\altaffiltext{2}{Institute of Physics and Astronomy, University of Potsdam, 14476 Potsdam, Germany}
\altaffiltext{3}{DESY, 15738 Zeuthen, Germany}

\begin{abstract}
\mpon{We perform particle-in-cell simulations of perpendicular nonrelativistic collisionless shocks  to study electron heating and pre-acceleration for parameters that permit extrapolation to the conditions at young supernova remnants. Our high-resolution large-scale numerical experiments sample a representative portion of the shock surface and demonstrate that the efficiency of electron injection is strongly modulated with the phase of the shock reformation. For plasmas  with low and moderate temperature (plasma beta $\beta_{\rm p}=5\cdot 10^{-4}$ and $\beta_{\rm p}=0.5$), we explore the nonlinear shock structure and electron pre-acceleration for various orientations of the large-scale magnetic field with respect to the simulation plane while keeping it at $90^\circ$ to the shock normal. Ion reflection off the shock leads to the formation of magnetic filaments in the shock ramp, resulting from Weibel-type instabilities, and electrostatic Buneman modes in the shock foot. In all cases under study, the latter provides first-stage electron energization through the shock-surfing acceleration (SSA) mechanism. The subsequent energization strongly depends on the field orientation and proceeds through adiabatic or second-order Fermi acceleration processes for configurations with the out-of-plane and in-plane field components, respectively. \mponn{For strictly out-of-plane field the fraction of supra-thermal electrons is much higher than for other configurations, because only in this case the Buneman modes are fully captured by the 2D simulation grid.} Shocks in plasma with moderate $\beta_{\rm p}$ provide more efficient pre-acceleration. The relevance of our results to the physics of fully three-dimensional systems is discussed.}
%Particle heating and acceleration, and effects of spontaneous turbulent magnetic reconnection, at high Mach number perpendicular nonrelativistic collisionless shocks are studied with two-dimensional fully kinetic Particle-In-Cell (PIC) simulations. \mpo{The parameters of the simulations are chosen to permit extrapolation to the conditions at young supernova remnants.} These unprecedented high-resolution large-scale simulations sample a representative portion of the shock surface to fully account for time-dependent effects of the cyclic shock reformation. The physics of strong shocks is governed by ion reflection that leads to the formation of magnetic filaments in the shock ramp, resulting from Weibel-type instabilities, and to electrostatic Buneman modes in the shock foot. We discuss the nonlinear shock structure and particle energization processes with emphasis on the dynamics of electron heating and pre-acceleration needed for their injection into diffusive shock acceleration. \mpo{We explore the} physical and numerical parameter space to find conditions providing the most efficient acceleration.  The importance of turbulent magnetic reconnection processes is scrutinized. \mpo{The} relevance of our results to the physics of fully three-dimensional systems is discussed.
\end{abstract}

\keywords{acceleration of particles, instabilities, ISM:supernova remnants, methods:numerical, plasmas, shock waves}

\section{Introduction}\label{introduction}
%Shell-type supernova remnants (SNR) are ideal laboratories for the study of particle acceleration at nonrelativistic collisionless shocks. The micro-physics governing the coupling of energetic charged particles, amplified turbulent magnetic field, and colliding plasma flows, determines how suprathermal particles can be injected into Fermi-type shock acceleration processes,what is the level of cosmic-ray feedback and hence the nonlinearity of the system . 

%\ab{Under construction...}

Acceleration of charged particles is a key topic in astrophysical research. High-energy particles are found \mpon{at astrophysical and \jn{interplanetary} shocks, and} collisionless \jn{nonrelativistic} shocks of supernova remnants (SNRs) are \jn{widely} believed to be the sources of \jn{galactic} cosmic rays \jn{(CRs)} with energies up to the knee at $\sim 10^{15}$ eV. 
\jn{Detection of high-energy $\gamma$-ray emission from SNRs by satellite and ground-based observatories prove the energetic particle production in these sources, although it is generally unclear whether the observed nonthermal emission comes primarily from protons or electrons \citep{2007A&A...464..235A}.}  
%Recently, using High Energy Stereoscopic System (HESS) TeV gamma rays from SNR have been detected, that gives us evidence of cosmic rays presence with energies up to 100 TeV \citep{2007A&A...464..235A}.

\jn{A dominant particle acceleration mechanism assumed to operate at nonrelativistic shock waves is}
%widely accepted particle acceleration theory assumes that dominant acceleration processes are 
diffusive shock acceleration (DSA), a first-order Fermi \jn{process} \citep[e.g.,][]{1983RPPh...46..973D,1987PhR...154....1B}. 
\jn{In this process particles gain their energies in repetitive interactions with the shock front. They are confined to the shock vicinity by magnetohydrodynamic (MHD) inhomogeneities that elastically scatter the particles in pitch-angle, providing for their diffusive motion in the upstream and downstream region of the shock. 
The most tantalizing unresolved question in DSA theory is \jn{the so-called particle} injection problem: DSA works only for particles whose Larmor radius is larger than the width of the shock transition layer, which is typically of the order of several proton gyroradii.} \mpo{Some pre-acceleration is thus required}, \jn{in particular} for electrons, \mpo{on account} of \jn{their lower} mass and \jn{consequently} smaller Larmor radii and inertial lengths, \jn{compared to protons}. 
%Thus, if we expect that DSA works some preacceleration mechanisms are needed to transfer electrons from thermal bulk to suprathermal and explain existence of nonthermal electron population in SNRs.

\jn{Here we study perpendicular shocks in a regime of high \jnn{Alfv\'enic and sonic} Mach numbers, $M_\mathrm{A} \gtrsim 30$ and $M_\mathrm{s} \gtrsim 50$, as appropriate for forward shocks of young SNRs.} \mpo{Using hybrid simulations, ion pre-acceleration has been shown to be inefficient at perpendicular shocks \citep[e.g.][]{2014ApJ...783...91C}, and we focus on electron pre-acceleration. The physics of such shocks is govered by ion reflection at the shock ramp,} and a variety of instabilities can be excited in the foot region due to interaction of upstream plasma with reflected ions \citep{2009A&ARv..17..409T}. The electrostatic two-stream instability, \mpo{also known as} Buneman instability \citep{Buneman1958}, is a result of interaction \mpo{between cold incoming} electrons and reflected ions. Early 1D simulations \citep{2002ApJ...572..880H} showed that the Buneman instability plays an important role in the generation of suprathermal electrons via shock surfing acceleration (SSA). Strong electrostatic waves are observed as a result of the nonlinear evolution of the Buneman instability.
While electrons are captured \mpo{in the electrostatic potential wells, they can be accelerated in perpendicular direction \jnn{by the convective electric field}. Multiple rapid interactions of electrons with electrostatic waves in the foot region of a shock give rise to} SSA and produce suprathermal tails in the electron spectra. \mpo{Two-dimensional simulations of
SSA at perpendicular shocks have been discussed in a number of papers \citep[e.g.,][]{2009ApJ...690..244A,2012ApJ...755..109M,2013PhRvL.111u5003M,2016ApJ...820...62W}. The trajectories of accelerated electrons in multidimensional systems are more complicated than in 1D, but the process is still referred to as SSA.
Our understanding of the efficiency of SSA and its dependence on the ion-to-electron mass ratio, the plasma temperature, the magnetic-field orientation with respect to the simulation plane, etc., is still incomplete.}

%Recent investigations of perpendicular shocks demonstrate variety of electron preacceleration processes in a foot region depending on mainly Alfvenic Mach number ($M_A$): for low $M_A$ regime quasi-standing whistler waves in foot region \citep{2011ApJ...733...63R}, shock drift acceleration (SDA) and first-order Fermi acceleration \citep{2014ApJ...794..153G}, sometimes shock surfing acceleration (SSA) \citep{2008ApJ...681L..85U,2009ApJ...695..574U} or even acceleration by magnetic reconnection events \citep{2015Sci...347..974M}; for high $M_A$ regime  mainly SSA \citep{2000ApJ...543L..67S,2012ApJ...755..109M,2013PhRvL.111u5003M,2010ApJ...721..828K}. In specific cases of high $M_A$ shocks growing of electrostatic Buneman instability is observed due to velocity difference between incoming (upstream) electrons and reflected ions  \citep{1988Ap&SS.144..535P,1988ApJ...329L..29C} and electrons could be trapped and accelerated by large-amplitude electric field. 
Since the Buneman instability is an electron-scale phenomenon, electron scales need to be resolved in simulations. \mpo{We conduct particle-in-cell (PIC) simulations in 2D3V configuration, meaning that we follow two spatial coordinates and all three components of velocity and electromagnetic fields. In contrast to hybrid simulations, PIC simulations follow electron trajectories as well as the ion dynamics.}

Our simulations complement previous 2D shock investigations \citep[e.g.,][]{2012ApJ...755..109M,2013PhRvL.111u5003M,2015Sci...347..974M,2016ApJ...820...62W}.
To explore electron energization via SSA, \mpo{the Alfv\'enic Mach number, $M_\mathrm{A}$, should satisfy a threshold condition for the excitation of electrostatic waves in the shock foot \citep{2012ApJ...755..109M}. The thermal velocity of the electrons should be smaller than the relative speed between incoming electrons and reflected ions, leading to \citep{2016ApJ...820...62W}}:
\begin{equation}
   M_s \gtrsim \frac{1+\alpha}{2} \ \sqrt{\frac{m_i}{m_e}} \ \sqrt{\frac{T_e}{T_i}},
\label{unstable}   
\end{equation}
where $\alpha\simeq 0.2$ \mpo{is the density ratio of reflected to incoming ions. Note that the ion temperature, $T_i$, is measured far upstream, whereas the electron temperature, $T_e$, is local, because the threshold condition involves the local thermal speed of electrons. \mponn{If the electron gyrofrequency is substantially lower than the plasma frequency ($\Omega_\mathrm{e}\simeq 0.1\ \omega_\mathrm{pe}$ in our simulations), the growth rate and wavevector of the Buneman mode are only weakly affected by the magnetic field, even if that is oriented perpendicular to the wavevector \citep{1982JGR....87..833S,Bohata}.}

In addition to the instability condition for the Buneman modes, the electrostatic} potential should be strong enough to trap electrons and hold them during acceleration. \mponn{\citet{2012ApJ...755..109M} use an estimate of the energy that is transferred from the incoming electrons to Buneman waves to find the balance between the \mpon{trapping force of the Buneman waves at} saturation level and the Lorentz force \jnn{for escaping}, which leads to the relation:}
\begin{equation}
  M_A \geq (1+\alpha)  \left(\frac{m_\mathrm{i}}{m_\mathrm{e}}\right)^{\frac{2}{3}}\simeq 26 ,
  \label{trapping}
\end{equation}
\mpo{where the last expression is derived for $\alpha=0.2$ and the mass ratio, \jnn{$m_i/m_e=100$}, used in our simulations.}

\mponn{Our choice of Alfv\'enic Mach numbers, $M_A \geq 30$,
%and and a low plasma beta $\beta \leq 0.5$ 
should therefore in all cases} lead to the formation of 
%Weibel-mediated 
shocks with strong electrostatic Buneman waves in the foot region. \jnn{However,} we note that even a moderate variation in the simulation parameters (e.g., $M_A$, $\beta_{\rm p}$, magnetic-field configuration) \jnn{may} introduce significantly different results: from absence of a nonthermal population \citep{2016ApJ...820...62W} to a large nonthermal fraction produced by the Buneman instability and adiabatic heating \citep{2012ApJ...755..109M}. Here we investigate the impact of the magnetic-field configuration, namely, the angle between the regular magnetic field and the simulation plane. We expect that these simulations may give us better understanding of acceleration processes in high-Mach-number shocks \jnn{under fully three-dimensional geometry}.

Early 1D simulations \citep{1985PhRvL..54.1872Q,1987PhFl...30.1767L} and recent 2D simulations \citep{2008ApJ...681L..85U,2009ApJ...695..574U,2014PhPl...21b2102U,2016ApJ...820...62W} indicated that supercritical perpendicular shocks become nonstationary and undergo cyclic self-reformation. \mpo{Therefore all instabilities driven by reflected ions in the foot region will also evolve in a nonstationary way, and the conditions for the Buneman-wave growth and efficient SSA will probably be met only at certain locations and time periods.} \jnn{Here} we follow the shock evolution \jnn{long enough (i.e.,} for 8 ion cyclotron times) 
%which is enough 
to demonstrate this influence.

The paper is organized as follows. We present description of simulations setup in Section~\ref{setup}. The results are presented in section~\ref{results}. Discussion and summary are in section~\ref{summary}.

\section{Simulation Setup} \label{setup}

\begin{figure}[htb]
\centering
\includegraphics[width=\linewidth]{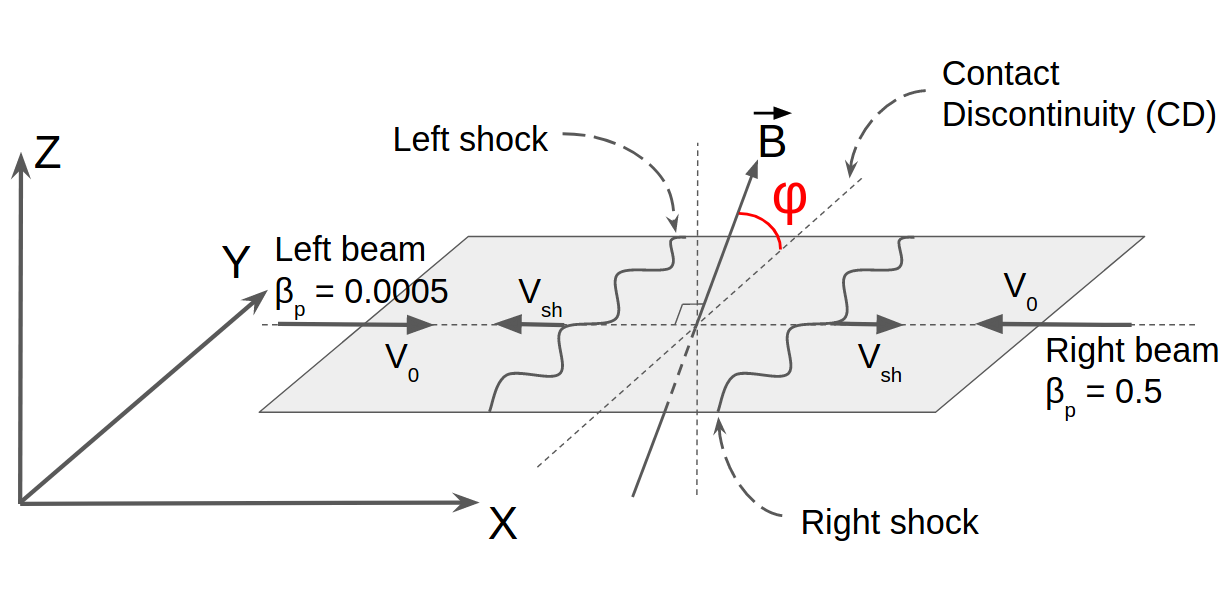}
\caption{Illustration of the simulation setup.} 
\label{setup1}
\end{figure}

\jn{In our simulations, two counter-streaming electron-ion plasma beams of equal density collide with each other to form a system of two shocks propagating in opposite directions \mpo{that are} separated by a contact discontinuity (CD; see Fig.~\ref{setup1}). The plasma flow is aligned with the $x$-direction, and the streaming velocities of the two slabs are $\boldsymbol{v}_{\rm L}=v_{\rm L}\hat{\boldsymbol{x}}$ and 
$\boldsymbol{v}_{\rm R}=v_{\rm R}\hat{\boldsymbol{x}}$, where the indices L and R refer, respectively, to the \emph{left} and \emph{right} sides of the simulation box, where the beams are injected.
The two beams carry a homogeneous magnetic field, $\boldsymbol{B}$, that is perpendicular to the flow direction and lies in the $yz$ plane, forming an angle $\varphi$ with the $y$-axis. As the magnetic field is assumed to be frozen into the moving plasma, a motional electric field $\boldsymbol{E}=-\boldsymbol{v}\times\boldsymbol{B}$ is also initialized in the \emph{left} and \emph{right} beam, with 
$\boldsymbol{v}=\boldsymbol{v}_{\rm L}$ or $\boldsymbol{v}=\boldsymbol{v}_{\rm R}$, respectively. 
The magnetic field strength in both plasmas is equal, $\boldsymbol{B}_{\rm L}=\boldsymbol{B}_{\rm R}$, and since $\boldsymbol{v}_{\rm L}=-\boldsymbol{v}_{\rm R}$, the motional electric field has opposing signs in the two slabs.  
To avoid an artificial electromagnetic transient resulting from this strong gradient in the motional electric field when the two plasma beams start to interact, we use a modified flow-flow method of shock excitation, recently developed by us and described in \citet{2016ApJ...820...62W}. The method implements a transition zone between the plasma beams, in which the electromagnetic fields are tapered off until they vanish in a small plasma-free area, that initially separates the beams. \mpo{Stability is provided by a current sheet that compensates $\boldsymbol{\nabla}\times\boldsymbol{B} $ in the transition layer.} In addition to providing a clean initialization setup, our method allows one to assume different physical conditions in the colliding beams, e.g., the asymmetry in the density of the slabs, as applied in \citet{2016ApJ...820...62W}.} 

\jn{Here we assume different temperatures for the left and the right plasma beam, that otherwise have the same physical characteristics\mpo{, including the density}. Specifically, we set the plasma beta (the ratio of the plasma pressure to the magnetic pressure) in the \emph{left} slab to $\beta_{\rm p,L}=0.0005$ and in the \emph{right} slab to $\beta_{\rm p,R}=0.5$.  The thermal velocities of plasma particles in the two beams thus differ by a factor of $\sqrt{1000} \simeq 30$, and so is the difference in the \emph{sonic} Mach numbers, $M_{\rm S}$, of the shocks that form on both sides of the CD. Note, that our choice of plasma beta $\beta_{\rm p}=0.5$ for one of the shocks allows for a direct \mpo{comparison} with the results of \citet{2012ApJ...755..109M} and \citet{2013PhRvL.111u5003M}.}
%CHECK THIS!

\jn{The counter-streaming plasma beams move with equal absolute velocities, $v_{\rm L}=0.2c=v_{\rm R}$, so they collide with a relative velocity of $v_{\rm rel}\simeq 0.38c$, where $c$ is the speed of light.
%NOTE ON SNR CONNECTION
Our simulation frame is the center-of-momentum frame of the system. Upon plasma collision, two shocks form \mpo{and} propagate away from the CD in the left and the right plasma. Here we refer to these shocks as to the \emph{left} and the \emph{right} shock, respectively. Because the \mpo{two unshocked plasmas are cold}, the system remains in approximate ram-pressure balance throughout the simulation, and the CD is stationary in the simulation frame. Therefore the simulation frame is coincident with the \emph{downstream} rest frames of the two shocks.} 

\jn{Restricted by available computational resources, we perform our simulations using a 2D3V model, i.e., we keep track of all three components of particle velocities and electromagnetic fields, and \mpo{follow particle positions only in the} $xy$ plane.
Since, as we show, in such a geometry the physics depends on the orientation of the initially uniform \emph{perpendicular} magnetic field with respect to the simulation plane, we carry out numerical experiments for three values of the angle $\varphi$, namely \emph{in-plane} magnetic field $\varphi=0^o$ (runs A1 and A2), $\varphi=45^o$ (runs B1 and B2), and \emph{out-of-plane} magnetic field $\varphi=90^o$ (runs C1 and C2; \jnn{see Fig.~\ref{setup1}}). Run-specific parameters are listed in Table~\ref{table-param}. Note that \mpo{the digits in the run designations above refer to the left and right shock, respectively, that formed in plasma} with different temperatures. They are listed here as separate runs to ease a comparison between the \mpo{cases of} \emph{cold} ($\beta_{\rm p}=5\cdot 10^{-4}$; runs A1, B1, C1) and \emph{moderate} ($\beta_{\rm p}=0.5$; runs A2, B2, C2) plasma beta.}

\jn{As noted, the magnetic field in both plasma beams is initially equal, $\boldsymbol{B}=\boldsymbol{B}_{\rm 0}$. We consider a weakly magnetized plasmas, and 
the ratio of the electron plasma frequency, $\omega_{\rm pe}=\sqrt{e^2N_e/\epsilon_0m_e}$, to the electron gyrofrequency, $\Omega_{\rm e}=eB_0/m_e$, is fixed to $\omega_{\rm pe}/\Omega_{\rm e}=12$. Here, $e$ and $m_e$ are the electron charge and mass, $\epsilon_0$ is the vacuum permittivity, and $N_e$ is the electron density. 
This value of plasma magnetization has been chosen in order to satisfy the trapping condition of the electron in the Buneman waves excited in the shock foot (see Section~\ref{introduction}).
We further assume the ion-to-electron mass ratio of $m_i/m_e=100$ and use 20 particles per cell per particle species for both plasma slabs. With this choice, the Alfv\'en velocity is numerically 
$v_{\rm A}=[B_{\rm 0}^2/\mu_{\rm 0}(N_em_e+N_im_i)]^{1/2}=8.29\times10^{-3}c$, where $\mu_{\rm 0}$ is the vacuum permeability.} 

\jn{The sonic and Alfv\'enic Mach numbers of the \jnn{two} shocks depend on the orientation angle of the uniform magnetic field \mpo{with respect to the simulation plane, $\varphi$}. For configurations with $\varphi=0^o$ and $\varphi=45^o$ (runs A and B), the large-scale field bends particle trajectories out of the simulation plane. Particles thus effectively have three degrees of freedom, \mpo{hence a} non-relativistic adiabatic index $\Gamma=5/3$. For the out-of-plane field configuration ($\varphi=90^o$, run C), particles are tied to the 2D simulation plane, have two degrees of freedom, and $\Gamma=2$. The resulting sound speeds thus differ, and for low plasma beta ($\beta_{\rm p}=5\times 10^{-4}$) they read $c_{\rm s}=(\Gamma kT_{\rm i}/m_i)^{1/2}\simeq 1.7\times10^{-4}c$ in runs A1 and B1, and $c_{\rm s}\simeq 1.86\times10^{-4}c$ in run C1. For a moderate plasma beta ($\beta_{\rm p}=0.5$; runs A2-C2), the sound speeds are a factor of $\sqrt{1000}$  \mpo{larger}. The compression \mpo{ratio} at the shock also depends on $\Gamma$ and is $r=3.97$ and $r=2.98$ for runs A-B and C, respectively.
%$c_{\rm s}\simeq 5.37\times10^{-3}c$ and $c_{\rm s}\simeq 5.88\times10^{-3}c$, respectively for runs A2-B2 and C2.  
The expected shock \mpo{speeds} in the \emph{simulation} frame are $v_{\rm sh}\simeq 0.067c$ for runs A and B, and $v_{\rm sh}\simeq 0.1c$ for run C. The \mpo{shock speeds} in the \emph{upstream} frame are $0.263c$ and $0.294c$, respectively. We calculate the sonic and Alfv\'enic Mach numbers of the shocks in the upstream reference frame, and their values are provided in Table~\ref{table-param}. Note, that for the parameters assumed in our simulations the shocks easily satisfy both the unstable ($M_{\rm s}\geq 6$) and the trapping condition ($M_{\rm A}\ga 25.8$), where our estimate uses $\alpha=0.2$ in Eqs.~\ref{unstable} and \ref{trapping}.
Thus in all cases we should expect efficient electron acceleration.}
%(sonic Mach numbers for cold and hot plasma slabs are 4490 and 142) and $0.294c$ (sonic Mach numbers for cold and hot plasma slabs are 5470 and 173) for runs A/B and C accordingly.

%In our simulations, two electron-ion plasma streams with different temperatures collide with each other at relative speed $v_{rel}=0.38c$, both streams move with absolute speed $v_0=0.2c$. Left plasma slab presents cold beam with the ratio of the electron pressure to the magnetic pressure (plasma beta) $\beta_e=0.0005$, for right plasma slab $\beta_e=0.5$ (hot beam).  
%The injected plasma carries a uniform magnetic field, $\vec{B}_0$, which is perpendicular to the plasma flow (lies in $y-z$ plane) and forms an angle $\varphi$ with the $y$-axis. We carry out three simulations with different configurations of upstream magnetic field, namely, with $\varphi$ equals $0^o$ (run A), $45^o$ (run B) and $90^o$ (run C).
%The special setup \citep{2016ApJ...820...62W} was used to avoid artificial antenna effect during the  initial collision of two plasma slabs.

%After the collision of two slabs of plasma forward (in hot region) and reverse (in cold region) shocks are formed and separated by a contact discontinuity. Expected absolute values of shock speed in simulation reference frame are $v_{sh}=0.067c$ for simulations A/B and $v_{sh}=0.1c$ for simulation C, values of shock speed in upstream reference frame are $0.263c$ (sonic Mach numbers for cold and hot plasma slabs are 4490 and 142) and $0.294c$ (sonic Mach numbers for cold and hot plasma slabs are 5470 and 173) for runs A/B and C accordingly. In fact, these values depend of the non-relativistic adiabatic index $\Gamma$, which equals 2 for simulation C and 5/3 for simulations A and B, because for out-of-plane magnetic field configuration particles move only in $x-y$-plane, so there are two degrees of freedom, but in simulations with $45^o$ and in-plane magnetic field configuration trajectories of particles are bended in z direction, it means that particles have three degrees of freedom.  

The electron skin \jn{depth in the} upstream plasma is common in all \jn{simulations} runs and equals $\lambda_{\rm se}=c/\omega_{\rm pe}=20\Delta$, where $\Delta$ is the size of the grid cells. \jn{For the assumed mass ratio, the ion skin depth is $\lambda_{\rm si}=200\Delta$. Here we use $\lambda_{\rm si}$ as the unit of length.}
The time scale and all temporal dependencies are given in terms of the upstream ion Larmor frequency $\Omega_{\rm i}^{-1}=120\omega_{\rm pi}^{-1}=1200\omega_{\rm pe}^{-1}=100\Omega_{\rm e}^{-1}$, \jn{where $\Omega_{\rm i}=eB_0/m_i$}. The simulation time, $\jnn{t}=8\Omega_i^{-1}$, is \jn{chosen} to cover \jn{at least} 3 shock reformation cycles. \jnn{The time-step we use is $\delta t=1/40\,\omega_{\rm pe}^{-1}=1/{48000}\,\Omega_{\rm i}^{-1}$.}

\jn{The two plasma beams are composed of an equal number of ions and electrons, initialized at the same locations to ensure the initial charge-neutrality.}
Plasma \mpo{is continuously injected at} both sides of the simulation box. The injection layer moves away from the interaction region \jn{and is all the times kept} at \mpo{sufficient distance to contain all reflected particles and generated electromagnetic fields in the computational box. At the same time, the distance is close enough so that} the beam does not travel too long without any interaction, which \mpo{suppresses numerical grid-Cerenkov effects and saves} computational resources.
\jn{The simulation box thus expands during an experiment in the $x$-direction, and can reach a final size $L_x=250 \lambda_{\rm si}=50,000\Delta$ at the end of the run. Open boundary conditions are imposed in the $x$-direction.}

The transverse size of the simulation box is $L_y=24 \lambda_{\rm si}=4800\Delta$ for runs A and B, and $12 \lambda_{\rm si}$ for run~C. \jn{Periodic boundaries are applied in the $y$-direction. The \mpo{box} size for runs A and B is larger because in these cases we expect to observe} turbulent magnetic reconnection within the shock structure \citep{2015Sci...347..974M}, \mpo{proper investigation of which} requires appropriate statistics for magnetic filaments formed in the shock ramp. The transverse box size \jn{is significantly larger} \mpo{than those used in earlier studies} \citep[e.g.][]{2012ApJ...755..109M,2015Sci...347..974M, 2016ApJ...820...62W}. 
%(e.g, $(5-6.8) \lambda_{si}$ in \cite{2012ApJ...755..109M,2015Sci...347..974M}, $12.6 \lambda_{si}$ in \cite{2016ApJ...820...62W}). 

The code used in this study is a 2D3V-adapted and modified version of the relativistic electromagnetic particle code TRISTAN with Message Passing Interface-based parallelization \citep{tristan,niemiec_2008}.
\jn{The numerical model is essentially the same as that used in \citet{2016ApJ...820...62W}. A notable  addition is the possibility to follow} individual \jn{selected} particle trajectories, which allows us to \jn{study} particle acceleration processes \mpo{in detail}.   
\jn{Convergence studies with various particle-in-cell numbers, $N_{\rm ppc}=10-40$, and values of the reduced mass ratio, $m_i/m_e=50-400$, have been performed to verify that the essential physical processes are correctly reproduced \mpo{for the parameters used here}.}

   \begin{table}
      \caption[]{Parameters of the Simulations and Derived Shock Properties.}
         \label{table-param}
\centering
\begin{tabular}{lccccl}
\hline
\hline
\noalign{\smallskip}
Run   & $\varphi$ & $L_y (\lambda_{\rm si})$ & $M_{\rm A}$ & $M_{\rm s}$ &$\beta_{\rm p}$ \\
\noalign{\smallskip}
\hline
\noalign{\smallskip}
A1  & $0^o$   & $24$  &  31.7  & 1550 & 0.0005\\
A2  & $0^o$   & $24$  &  31.7  & 49   & 0.5\\
\noalign{\smallskip}
\hline
\noalign{\smallskip}
B1  & $45^o$  & $24$  &  31.7  & 1550 & 0.0005\\
B2  & $45^o$  & $24$  &  31.7  & 49   & 0.5\\
\noalign{\smallskip}
\hline
\noalign{\smallskip}
C1  & $90^o$  & $12$  &  35.5  & 1581 & 0.0005\\
C2  & $90^o$  & $12$  &  35.5  & 55   & 0.5\\
\noalign{\smallskip}
\hline
\end{tabular}
\smallskip
\tablecomments{Parameters of the simulation runs described
in this paper. Listed are: the orientation of the uniform perpendicular magnetic field with respect to the 2D simulation plane, $\varphi$, the transverse size of the computational box size in units of the ion skin depth, $\lambda_{\rm si}=200\Delta$, Alfv\'enic and sonic Mach numbers of the shocks, $M_{\rm A}$ and $M_{\rm s}$, and the plasma beta, $\beta_{\rm p}$, of the upstream plasmas. All runs use an electron skin depth of $\lambda_{\rm se}=20\Delta$, the ion-electron mass ratio $m_i/m_e=100$, and the plasma magnetization $\omega_{\rm pe}/\Omega_{\rm e}=12$.} 
   \end{table}

\section{Results} \label{results}

In what follows, we present comparison of \mpo{three types of simulations differing in the orientation of the upstream magnetic field: in-plane ($\phi=0^o$, runs A1 and A2), $\phi=45^o$ (runs B1 and B2) and out-of-plane ($\phi=90^o$, runs C1 and C2). We discuss the overall shock structure, differences in the evolution of the Buneman modes, the influence of these modes on electron pre-\jnn{acceleration}, the main \jnn{features} of electron acceleration for the \jnn{three} magnetic-field configurations, \jnn{the resulting electron spectra downstream of the shock,} and the} influence of shock \jnn{self-}reformation on all these processes.

\subsection{Global Shock Structure}
In this section we \jn{give an overview of the} shock structures \mpo{observed for the three cases under study. We base our description on the \emph{left} shocks propagating into cold plasma, i.e., runs A1-C1, and discuss results for shocks developing in a moderate-temperature plasma, runs A2-C2, only where they differ from those for $\beta_\mathrm{p}\ll 1$.}

\begin{figure}[t!]
\centering
\includegraphics[width=\linewidth]{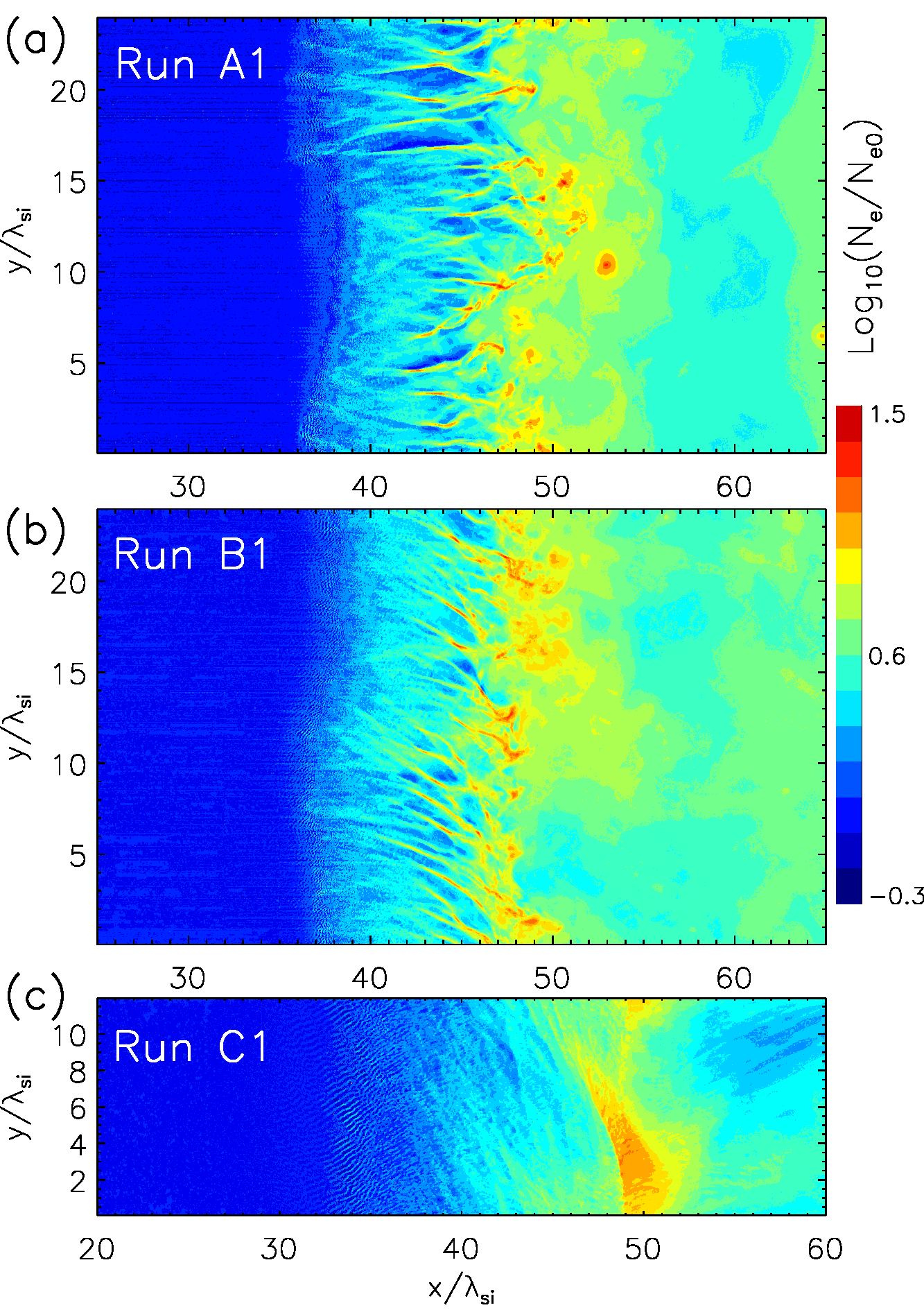}
\caption{\jn{Distributions of the normalized electron \jnn{number} density at shocks propagating in cold plasmas. Panel (a): run A1 at $t\Omega_{\rm i}=7.625$, panel (b): run B1 at $t\Omega_{\rm i}=7.5$, and panel (c): run C1 at $t\Omega_i=7.75$. A logarithmic scaling is used.} %Panel (a): upstream region ($x<34\lambda_{si}$), foot region ($34\lambda_{si}<x<48\lambda_{si}$), shock overshoot -  $50\lambda_{si}$ and downstream region ($x>50\lambda_{si}$). Panel (b): upstream region ($x<34\lambda_{si}$), foot region ($34\lambda_{si}<x<48\lambda_{si}$), shock overshoot -  $48\lambda_{si}$ and downstream region ($x>48\lambda_{si}$). Panel (c): upstream region ($x<32\lambda_{si}$), foot region ($32\lambda_{si}<x<47\lambda_{si}$), shock overshoot -  $49\lambda_{si}$ and downstream region ($x>49\lambda_{si}$)
} 
\label{density}
\end{figure}

\begin{figure*}[t!]
\centering
\includegraphics[width=\linewidth]{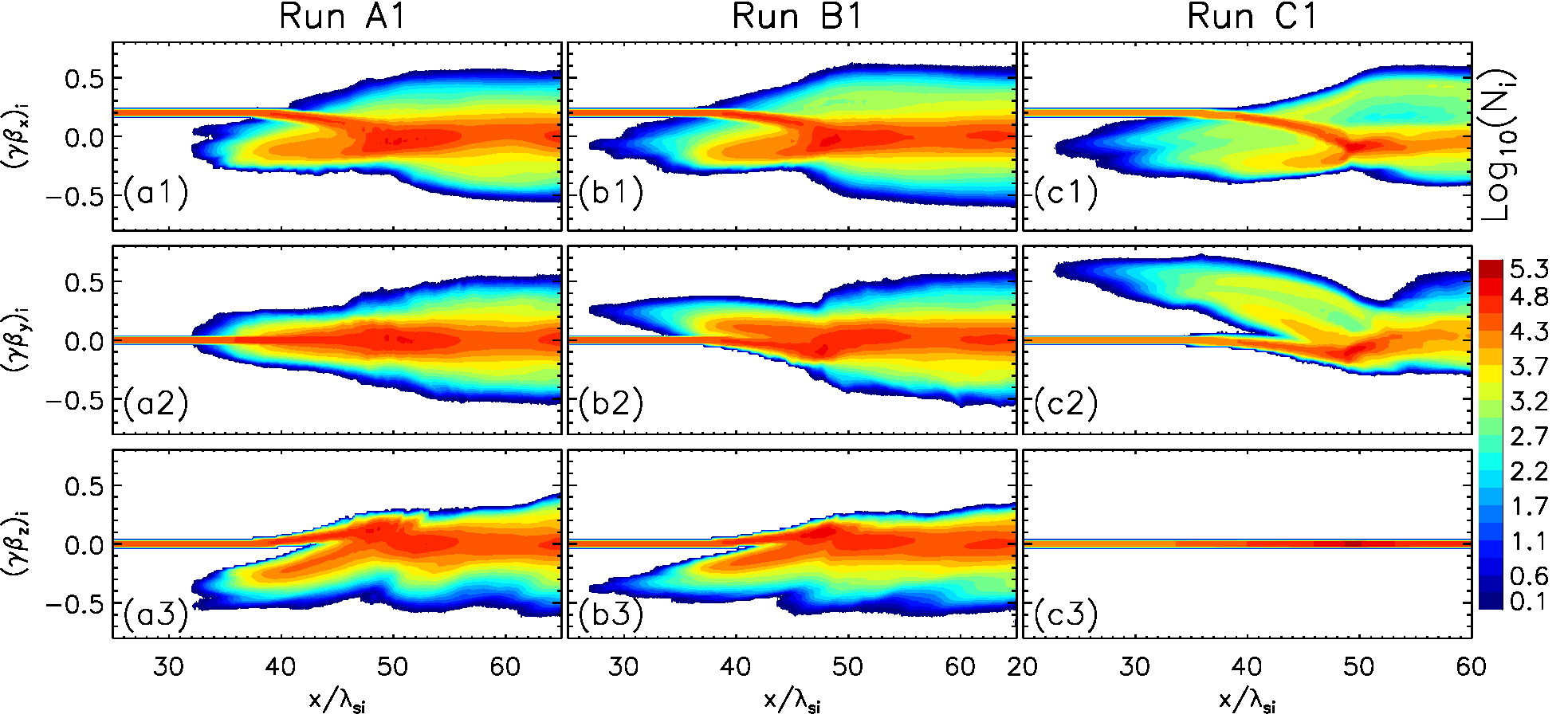}
\caption{Phase-space distributions of ions \jn{for the \emph{left} shock regions shown in Fig.~\ref{density}, \mpo{averaged \jnn{over} the spatial coordinate $y$}.}
From top to bottom, \jn{shown are the} $x$, $y$, and $z$-components of \jn{particle momenta,} $\gamma\beta$, \jn{from left to right, for} run A1, B1, and C1. A logarithmic scaling is used.} 
\label{phase_space}
\end{figure*}

Figures~\ref{density} and \ref{phase_space} show electron density maps \jn{and the ion phase-space} in the shock region \jn{in panels (a), (b), and (c), respectively, for runs A1, B1, and C1. They present the system at times} close to the end of the simulation runs \jn{and specifically at a phase of shock reformation in which the largest number of shock-reflected ions appear, and consequently the Buneman waves in the shock foot reach maximum amplitudes. These times differ slightly between the runs and are} $t\Omega_{\rm i}=7.625$ for run A1, $t\Omega_{\rm i}=7.5$ for run B1, and $t\Omega_i=7.75$ for run C1. 
\jn{Note that the system already contains fully-formed self-sustained shocks.}

\begin{figure}[htb]
\centering
\includegraphics[width=\linewidth]{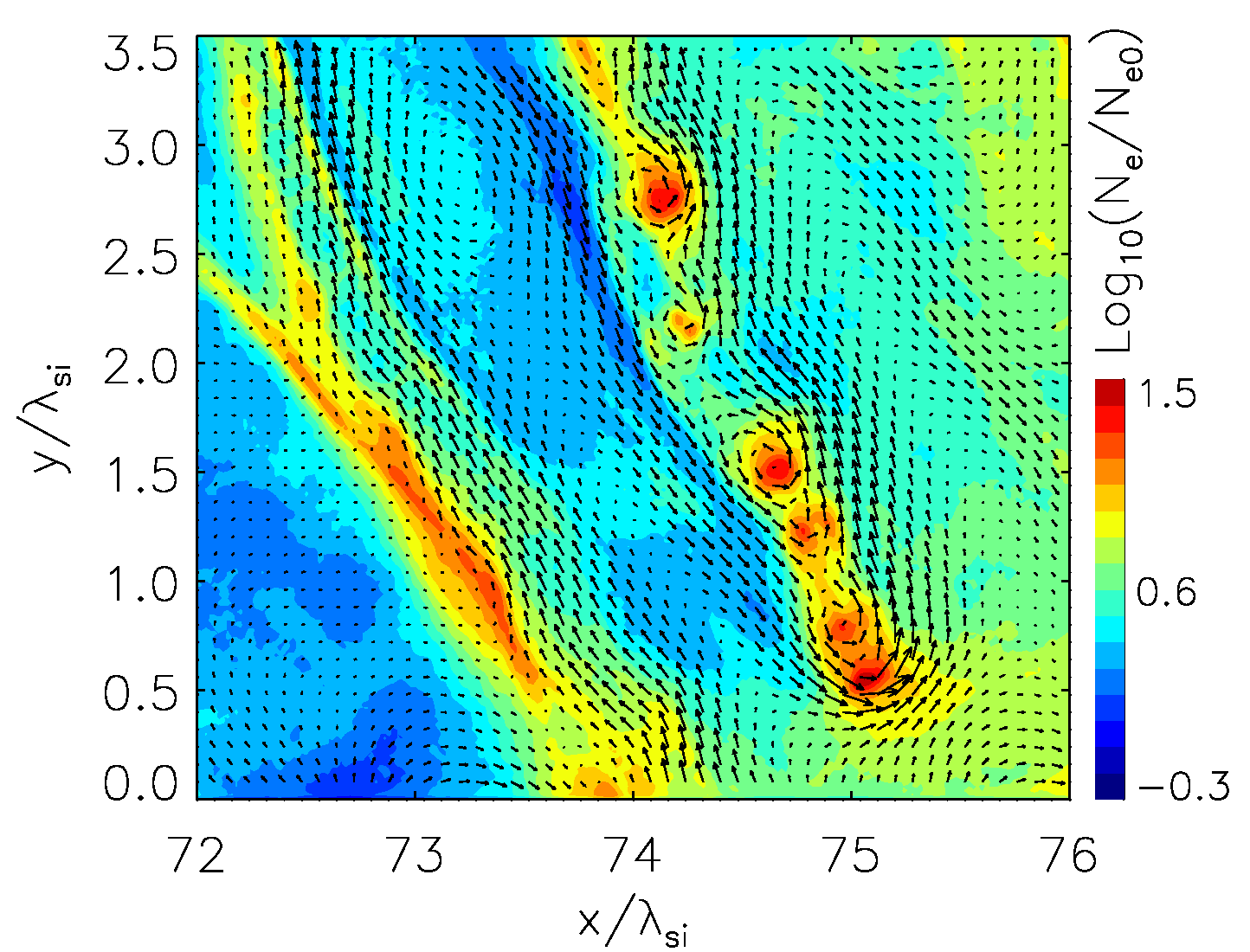}
\caption{Blow-up of the region \jn{containing} a chain of magnetic islands \jn{resulting from magnetic reconnection in the filamentary structures of the shock ramp.} The snapshot was taken for run B1 at time $t=3.8\Omega^{-1}$. \mpo{The normalized density is color-coded in logarithmic scale}. Arrows show the in-plane ($x-y$) component of the magnetic field.} 
\label{reconnection-region}
\end{figure}

\jn{As expected for} supercritical shocks, \jn{their structures are determined by} the fraction of upstream \jn{plasma} ions that are reflected from the shock front. \jn{In our case of perpendicular shocks, the reflection is due to the shock-compressed magnetic field. Reflected ions} gyrate \jn{around the magnetic-field lines} in the upstream region, \jn{exciting various plasma instabilities. \mpo{For shocks with} high Alfv\'en Mach number, the most important instabilities are the Weibel-type filamentation instability in the shock ramp and the Buneman instability in the shock foot \citep{2016ApJ...820...62W}.
\mpo{Ion reflection also leads} to the so-called overshoot, i.e., plasma compression at the shock front that exceeds the compression expected from the Rankine-Hugoniot conditions in the MHD description. 
The overshoot in run C1 with the out-of-plane uniform magnetic field can be approximately identified with a largely-coherent compression structure at $x/\lambda_{\rm si}\approx 50$ in Figure~\ref{density}c (compare also Fig.~\ref{phase_space}). In cases A1 and B1, the shock transition does not produce a coherent structure, but in the density profiles averaged over the $y$-direction (not shown) the overshoot is located at $x/\lambda_{\rm si}\approx 50$ and  $x/\lambda_{\rm si}\approx 48$, respectively, for runs A1 and B1. Downstream of the overshoot the plasma density oscillates around an average value \mpo{that is commensurate with that expected in the MHD picture} (see Sections~\ref{introduction} and \ref{reformation-nonthermal}). } 

\jn{The Weibel-type filamentation instability results from the}
interaction \mpo{between} \jn{shock-}reflected and incoming plasma ions. It leads \jn{to the formation of magnetic filaments whose separation scale is of the order of the} ion skin \jn{depth, $\lambda_{\rm si}$. 
The filaments can be identified} in the density distributions \jn{(Fig.~\ref{density}) in the shock-ramp region between the overshoot and the shock foot, i.e., for $x/\lambda_{\rm si}\approx (38-50), (38-48)$, and $(36-50)$, for runs A1, B1, and C1.} The structure and geometry of filaments \mpo{depend on the} configuration of the uniform magnetic field, \mpo{because it defines the gyration} direction of the reflected ions. \mpo{Figure~\ref{phase_space} demonstrates that in} the case of an in-plane magnetic field \jn{($\varphi=0^o$, Fig.~\ref{phase_space}a) ions are reflected primarily in the $xz$ plane, which} leads to the formation of \mpo{the density filaments \jn{along the plasma flow direction} that we see in Figure~\ref{density}a}. 
For the configuration with $\varphi=45^o$ \jn{(Fig.~\ref{phase_space}b), in addition to} negative $v_x$ and $v_z$ components of \jn{the reflected ion} velocity, there is a positive $v_y$ velocity component \jn{that causes the density filaments to} become oblique. \mpo{Finally,} for the out-of-plane configuration
\jn{($\varphi=90^o$, Fig.~\ref{phase_space}c)} the reflected ions \jn{are confined to the simulation plane, and again density filaments result} \mpo{that are blurred in Figure~\ref{density} on account of their large obliquity.} 

\jn{The waves visible in the density distributions in Figure~\ref{density} upstream of the shock ramp, in the \emph{shock foot} regions ($x/\lambda_{\rm si}\approx (34-38)$ for runs A1 and B1, and $x/\lambda_{\rm si}\approx (30-36)$, for run C1),} have a different nature \jn{and result from the electrostatic} Buneman instability caused by the interaction between the shock-reflected ions and \jn{inflowing} upstream electrons. \jn{Their} wavelength is much smaller than the ion inertia length, and the wave vector is \jn{approximately} orthogonal to \mpo{that of the} magnetic filaments. The Buneman waves are discussed in detail in Section~\ref{Buneman}.

It was demonstrated by \citet{2015Sci...347..974M} for the \emph{in-plane} magnetic field configuration, \jn{that merging} magnetic filaments \jn{in the shock ramp} can trigger spontaneous turbulent magnetic reconnection, providing an additional channel for electron acceleration. We observe magnetic reconnection \jn{for both the} in-plane and $\varphi=45^o$ \jn{magnetic field} configurations \jn{(runs A and B for the cold and moderate-temperature plasmas). The magnetic-reconnection events can be identified by chains of magnetic islands separated \mpo{by X-points, which result from} nonlinear decay of the current sheets \citep{1963PhFl....6..459F} %\jnn{PROVIDE REFERENCE TO MAYBE DRAKE?}. 
Corresponding \mpo{enhancements lined up along magnetic filaments are visible in plasma density, an example of which is shown in Fig.~\ref{reconnection-region}}. The role of the magnetic-reconnection specific acceleration processes will be discussed in detail in a follow-up paper.} 

\subsection{Cyclic Shock Reformation} \label{reform}

\begin{figure}[b!]
\centering
\includegraphics[width=\linewidth]{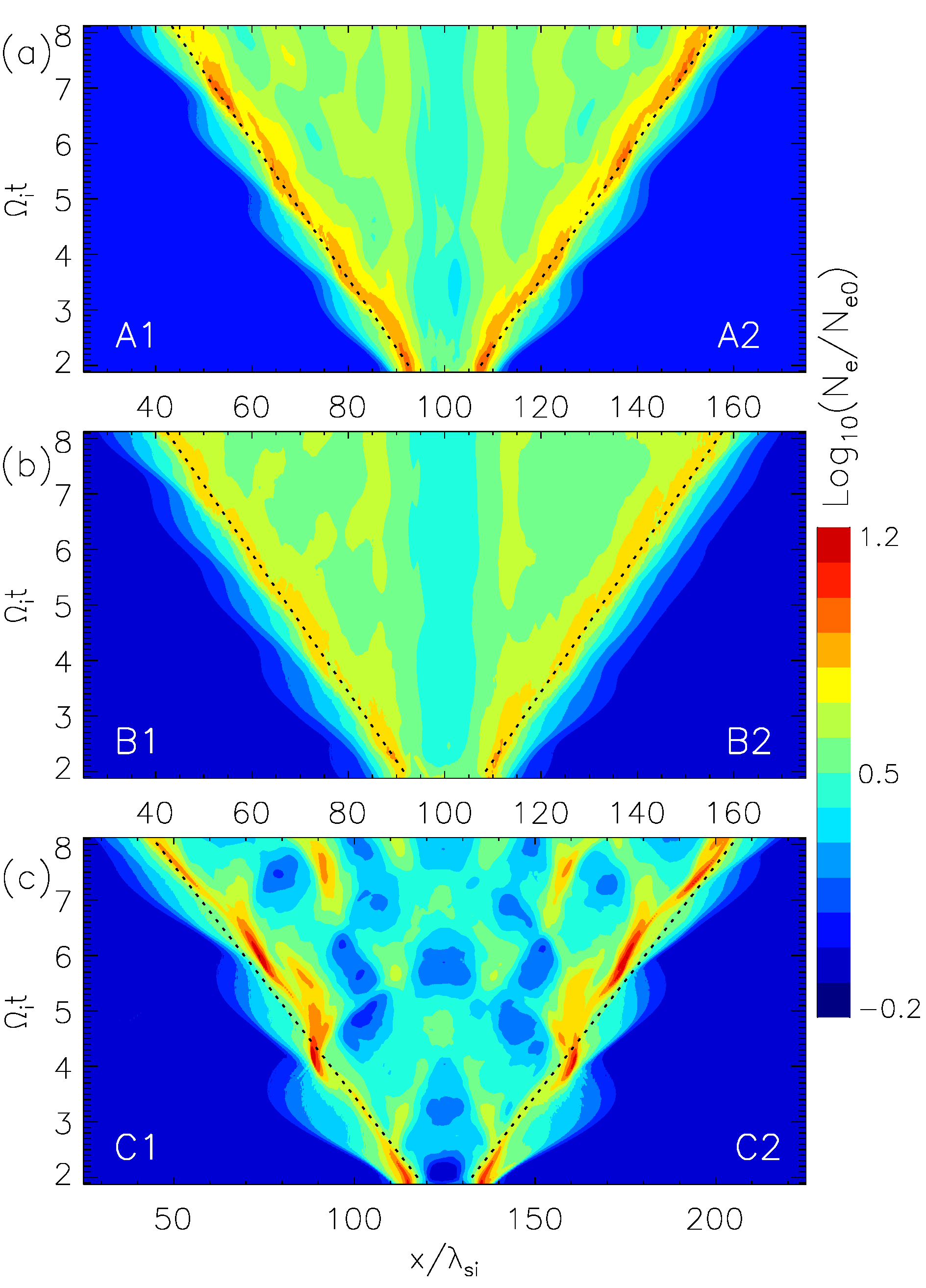}
\caption{Time development of the electron number density averaged over the $y$-direction for runs A1, A2 (a), runs B1, B2 (b), and runs C1, C2 (c). Color contours present normalized electron density in logarithmic scale. \mpo{The dashed lines indicate the \ab{theoretical} shock speed in the simulation frame.}}
\label{shock_reform}
\end{figure}

\jn{Our previous investigation of perpendicular shocks in the regime of high Alfv\'en Mach number \citep{2016ApJ...820...62W} demonstrated a cyclic shock self-reformation known \mpo{from earlier studies of} low-Mach-number shocks. 
The process is caused by non-stationary ion reflection off the shock ramp that at high Mach numbers was shown to be governed by the dynamics of magnetic filaments formed by the Weibel-type instability. The time-scale of shock reformation is on the order of the \mpo{gyroperiod of upstream ions}. During reformation the shock velocity and plasma density at the shock are quasi-cyclicly modulated around their average values, and the extension of the filamentary region in the shock ramp varies. A~bunch of shock-reflected ions streaming against incoming plasma leads to the formation of current filaments extended along the plasma flow direction \mpo{that are accompanied in the shock foot by} electrostatic Buneman modes. Later in the cycle, once the current filaments start to merge ahead of the shock ramp and become aligned closer to the shock surface \mpo{on account of} bunched ion gyration, the turbulent shock precursor shrinks, and the Buneman modes disappear. This has profound consequences for particle acceleration, as will be discussed in Section~\ref{reformation-nonthermal} below.}

Figure~\ref{shock_reform} shows for all runs the time evolution of the electron density averaged over the $y$-direction. \jn{The left and right shocks' motion with an average speed $v_{\rm sh,L}$ or $v_{\rm sh,R}$ are marked with dashed lines. A self-reformation cycle begins at $T\Omega_i\approx 2$, after
the shocks have been \jn{fully} formed. The reformation processes are observed for all magnetic-field orientations and plasma temperatures studied and are most pronounced \mpo{for} out-of-plane (run C) and in-plane (run A) field configurations.} The period of self-reformation varies \ab{around the average value $1.55\Omega_i^{-1}$} \jn{and is consistent with our earlier finding of approximately $1.5\Omega_i^{-1}$ obtained for the $\varphi=45^o$ configuration \citep{2016ApJ...820...62W}.} \mpo{Rippling modes caused by spatial modulation of ion reflection \citep[e.g.,][]{2007PhPl...14a2108B,2016ApJ...820...62W} are not captured in our simulations because their wavelength is not smaller than the transverse size of the simulation box. }

%For run B2 (fig.~\ref{shock_reform}(b), right shock) shock after $t=4\Omega_i^{-1}$ there is almost steady shock propagation with constant speed. 
%Such local stationarity of the shock propagation is caused by the stationarity of ion reflection from shock overshoot and probably by the shock rippling \citep{2016ApJ...820...62W}. It means that each small part of the shock front undergoes self-reformation processes and temporal evolution of density profile is similar to that shown on figure~\ref{shock_reform}(b)(left shock). Only run B2 demonstrates stationar shock propagation while in all other cases shock front undergoes cyclic self-reformation.

%Summarize our results and taking into account the results of \cite{2010ApJ...721..828K} (nonrelativistic perpendicular shock with in-plane magnetic field configuration and plasma beta $\beta=26$): 3 cases with clear self-reformation for $\beta=0.0005$, 2 cases with clear self-reformation and 1 case with steady shock development for $\beta=0.5$ and steady shock for $\beta=26$, we can mention about probable dependence between shock reformation processes and plasma beta, namely for higher $\beta$ the shock front moves in more regular way.

%The period of self-reformation varies in the range from $1.2\Omega_i^{-1}$ to $1.9\Omega_i^{-1}$.  Unfortunately we can not say how the period of self-reformation depends on magnetic field orientation or plasma temperature because of low reformation cycles statistics (only 3 cycles for each case) and wide range of period variations.

\mpon{The velocities of the \jnn{left} and the \jnn{right} shock in the simulation (or downstream) reference frame (calculated as an overshoot speed) vary between $0.03c$ and $0.15c$.} The average shock speed equals $0.066c$ for runs A1, A2, $0.067c$ for runs B1, B2 and $0.094c$ for runs C1, C2, \mpo{very close to the expected speeds of $0.066c$ and $0.1c$, respectively}.

\subsection{The Buneman Instability} \label{Buneman}
As discussed above, the Buneman instability results from the interaction of shock-reflected ions with the upstream electrons. It has been shown that this instability \jn{can be excited in the foot of high-Mach-number} nonrelativistic perpendicular shocks. \jn{The properties of \mpo{these electrostatic modes depend on} physical parameters such as the shock Mach number and plasma temperature. In addition, the instability characteristics may be different in numerical studies with restricted dimensionality. In particular, in 2D simulations the orientation of the average magnetic field with respect to the simulation plane may play a role, as suggested from a comparison of studies applying the out-of-plane magnetic field \citep[e.g.,][]{2012ApJ...755..109M,2013PhRvL.111u5003M}, the $\varphi=45^o$ configuration \citep[e.g.,][]{2016ApJ...820...62W}, and the in-plane field \citep[e.g.,][]{2010ApJ...721..828K}. However, all these works differ in the specific physical parameters assumed. In contrast, our present study enables a direct comparison of the Buneman instability properties in simulation runs in which} the \mpo{main} variable parameter is the magnetic field orientation.

\begin{figure}[htb]
\centering
\includegraphics[width=\linewidth]{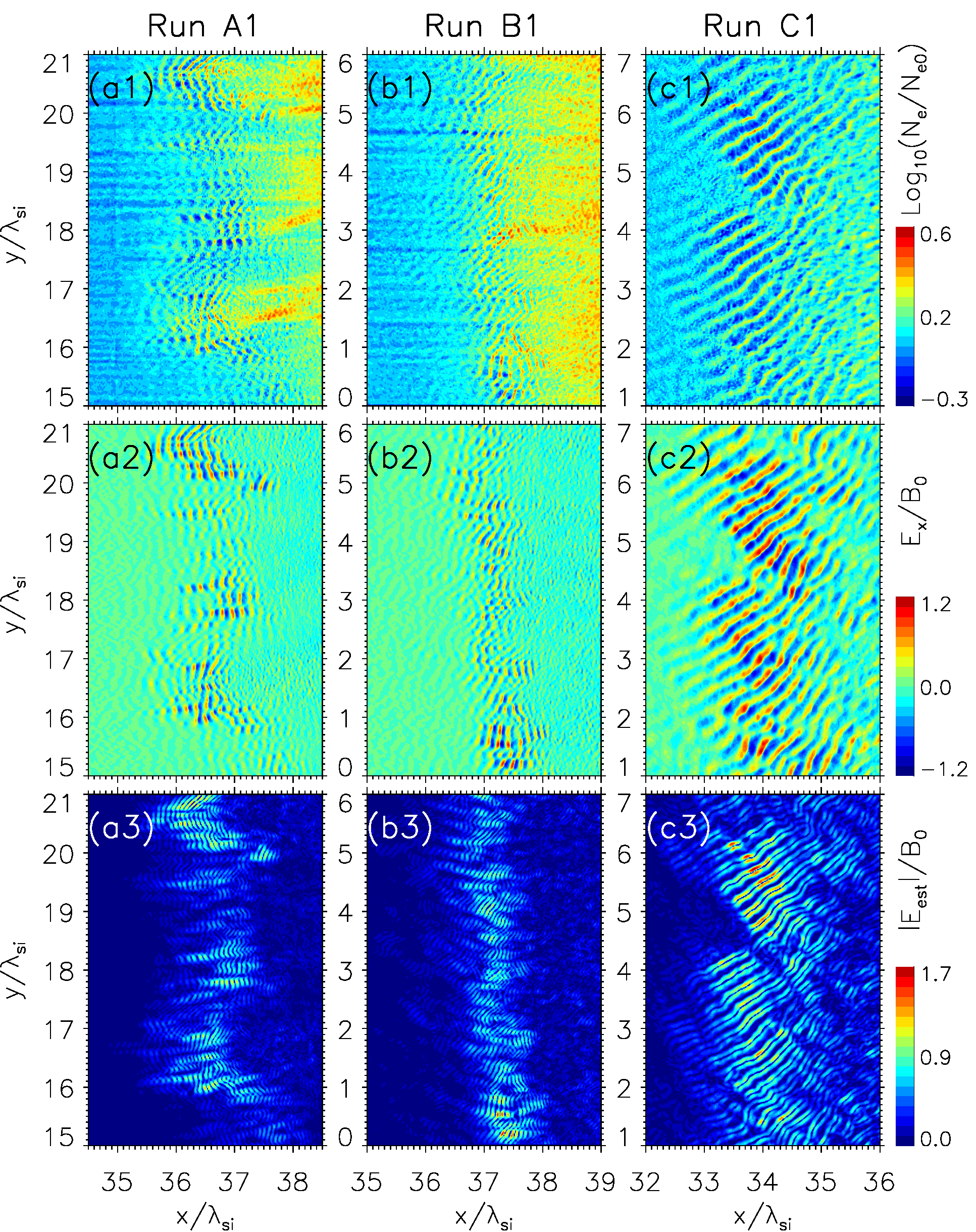}
\caption{\mpo{Electron density (a1, b1, c1), the $x$ component of the electric field (a2, b2, c2), and electrostatic field amplitudes (a3, b3, c3) in selected regions \jnn{of the shock foot} with the most intense Buneman waves for runs A1 (a* panels), B1 (b*), and C1 (c*). Density is presented in logarithmic scaling and normalized to the upstream density. Electric and electrostatic field strengths are normalized to that of the upstream magnetic field.}}
\label{Eest}
\end{figure}

%\begin{figure*}[t!]
%\centering
%\includegraphics[width=\linewidth]{four2D_Epar-2.eps}
%\caption{Fourier power spectrum of electric field parallel to the wave vektor, $e_k\cdot E$ for the regions presented on figure~\ref{Eest} runs A (a), B (b) and C (c).}
%\label{Fourier}
%\end{figure*}

%\begin{figure*}[t!]
%\centering
%\includegraphics[width=\linewidth]{four2D_Epar-3.eps}
%\caption{Fourier power spectrum of electric field parallel to the wave vektor, $e_k\cdot E$ for the regions presented on figure~\ref{Eest} runs A (a), B (b) and C (c).}
%\label{Fourier}
%\end{figure*}

\begin{figure*}[t!]
\centering
\includegraphics[width=\linewidth]{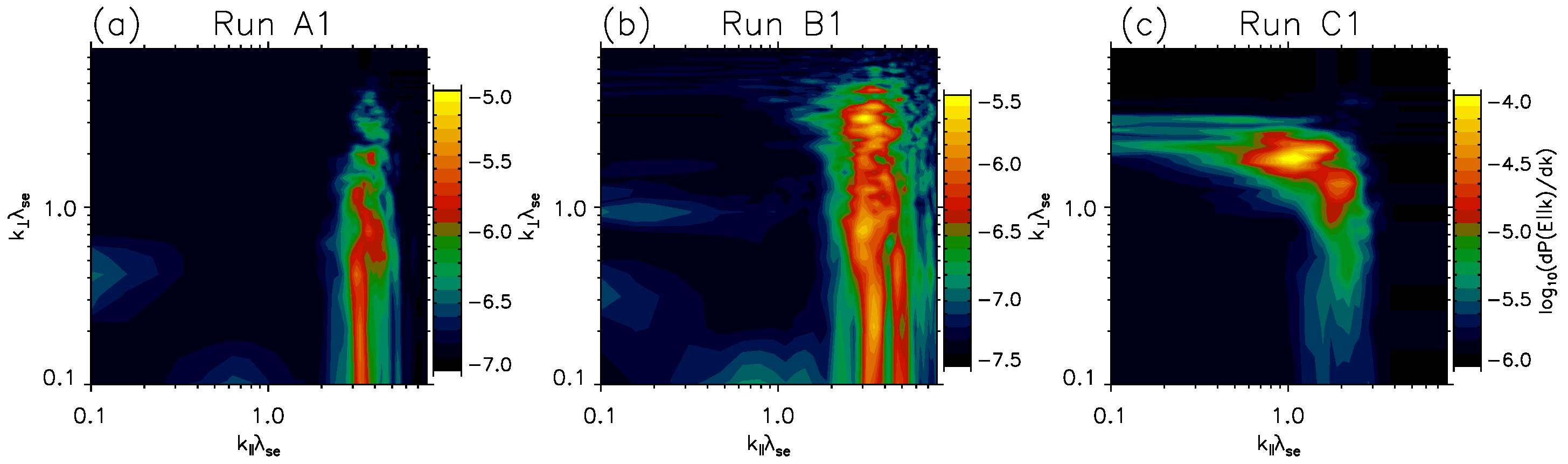}
\caption{Fourier power spectrum of \jnn{the} electric field parallel to the wave \mpo{vector, $\mathbf{e}_k\cdot \mathbf{E}$, for the regions presented in the panels a1 (a), b1 (b) and c1 (c) of Fig.~\ref{Eest}}.}
\label{Fourier}
\end{figure*}

\begin{figure}[htb]
\centering
\includegraphics[width=\linewidth]{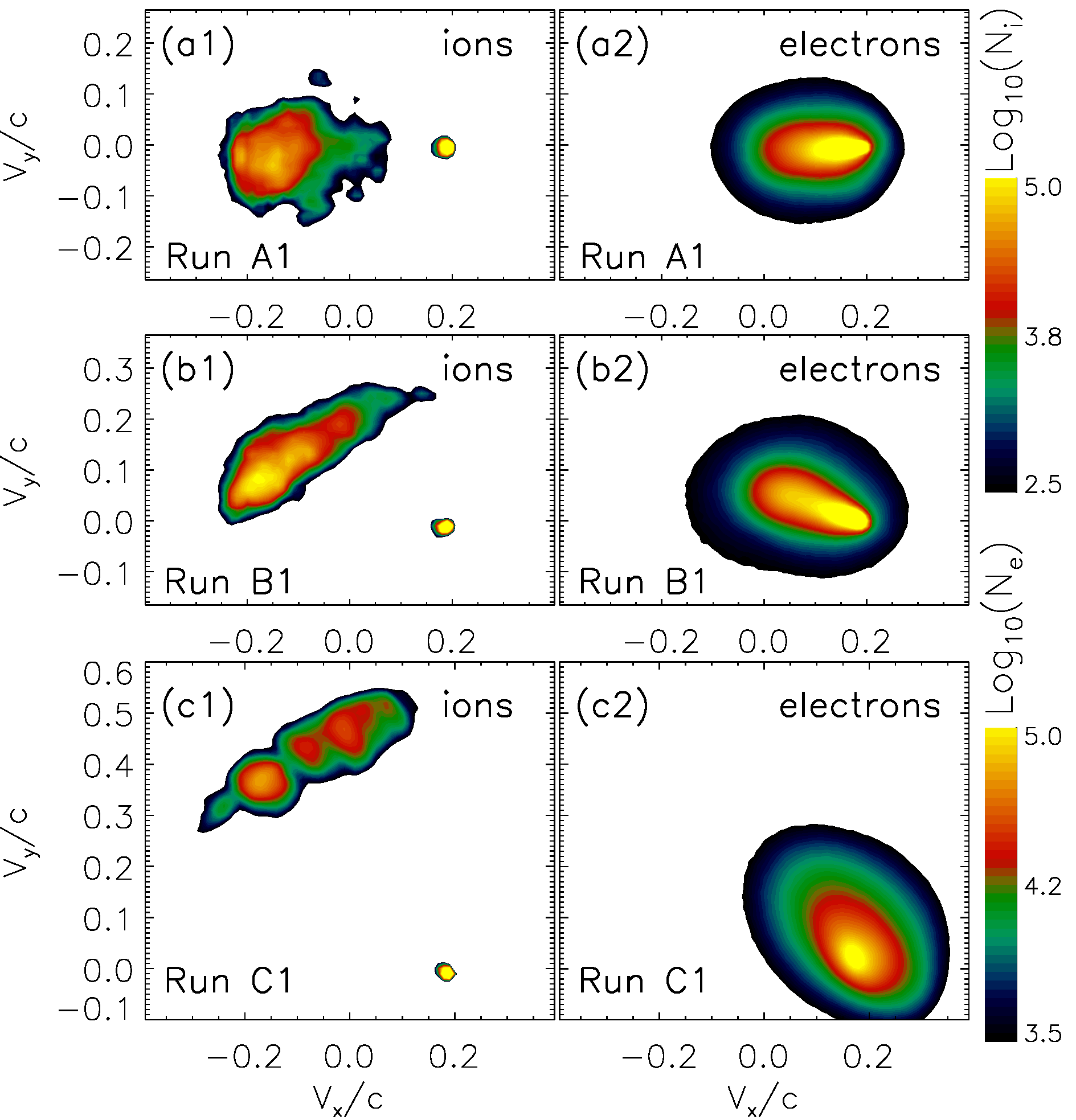}
\caption{Phase-space distribution in $v_x-v_y$ of ions (first column) and electrons (second column) \mpo{in the} regions presented in Fig.~\ref{Eest} for runs A1 (a1, a2), B1 (b1, b2) and C1 (c1, c2). The scale is logarithmic.}%\ab{Everything is ok with these plots. Panels (c1) and (c2) cover bigger range of velocities, that's why "cloud" of electrons was smaller (already fixed). And second thing: electron "clouds" in panels (a2) and (b2) have a shift in the direction to the point (0;0) (in the direction of smaller energies) while in panel (c2) there is a trend in almost opposite direction (bigger energies) and it explains high energy tail in a spectra plot.}}
\label{velocities}
\end{figure}

\jn{Figure~\ref{Eest} presents a \mpo{blow-up of a portion} of the \emph{left} shock foot shown in Figures~\ref{density} and \ref{phase_space}. From top to bottom, displayed are the}
electron density, the $E_x$ component of the electric field, and the electrostatic field amplitude, \jn{from left to right for runs A1, B1, and C1.} The electrostatic field \mpo{amplitude} is \jn{calculated as} $|E_{est}|=|-\nabla\phi|$, where $\phi$ is the electric potential derived directly from the charge distribution. \jn{The modes in the electrostatic field maps} \mpo{in Figures~\ref{Eest}a3-\ref{Eest}c3 appear to have half their true wavelength}, because the absolute values of $\nabla\phi$ \jn{are plotted}, \jn{and one cannot distinguish between positive and negative field \mpo{strengths}.
The wave structures visible in the plot occur at scales much smaller than the ion inertia scale. The form of the waves \mpo{depends on} the magnetic-field configuration. For the out-of-plane field orientation (run C1, Fig.~\ref{Eest}c) coherent wave trains are formed over an extended region of the shock foot \citep[see, e.g.,][]{2012ApJ...755..109M}.
The electrostatic field reaches amplitudes of $|E_{est}|/B_{\rm 0}\sim 2$, \mpo{and} the trapping condition is easily satisfied. \mpo{Wave coherence} is broken for the in-plane configuration (run A1, Fig.~\ref{Eest}a), and the modes are distributed in patchy structures formed in narrow regions ahead of the magnetic filaments (see Fig.~\ref{Eest}a1). In the $\varphi=45^o$ case, the \mpo{appearance is intermediate between these forms} -- the wave structure is largely coherent but over a narrow region with thickness $\sim\lambda_{\rm si}$. For both A1 and B1, the amplitude of the electrostatic field %{can still} 
reach $|E_{est}|/B_{\rm 0}\sim 1.7$ at \mpo{peak locations}, but the average field strength is $|E_{est}|/B_{\rm 0}\sim 1$ or less. 
%Electrostatic field strength averaged over regions presented in figure~\ref{Eest}(a3,b3,c3) equal 0.019, 0.021 and 0.034 per cell for runs A1, B1 and C1 respectively.

\mponn{A significant part of the variation in the coherence and the volume coverage between runs A/B and C arises from the homogeneity of the reflected-ion beam which in turn depends on the structure of the overshoot. Ion reflection in the out-of-plane case is largely uniform along the shock because the shock overshoot has a fairly coherent structure. The overshoot region for $0^o$ and $45^o$ magnetic-field orientations has clumpy structures producing an incoherent flow of reflected ions.}

As noted in Section~\ref{reform}, the intensity of the electrostatic waves varies considerably \mpo{during a shock reformation cycle on account of the changing number of} shock-reflected ions. The \ab{average value of the}
electrostatic field amplitude varies in the range $|E_{est}|/B_{\rm 0}\sim 0.1-1.1$  for run A1, $0.3-1.1$ for run B1, and $0.3-1.5$ for run C1. %{\jn{PROVIDE NUMBERS}}. 
\jn{Note also that for the right shocks in moderate-temperature plasma,} \jnn{$\beta_{\rm p}=0.5$, the maximum} \mpon{intensity of
the Buneman waves is} $20\%-50\%$ smaller \mpo{than in the cold plasma. \jnn{However, the} area of the unstable region is $20\%-30\%$ larger in \mpon{the case of} warm plasma. We do not expect a thermal reduction in the Buneman growth rate that would explain the lower wave intensity, and we \mponn{cannot exclude} that the saturation level is lower on account of strong nonlinear Landau damping \citep{breizman+ryutov} or the modulation instability \citep{1977JETP...45..266G}. The efficiency of electron acceleration is determined by the number of trapped electrons \jnn{and} the volume fraction occupied by intense Buneman modes \jnn{with} the convective electric field that provides the acceleration. \jnn{A conjunction of these factors, though different in each case analysed here,} permit efficient electron acceleration \jnn{at both the low- and moderate-$\beta_{\rm p}$} shocks.}

\mpo{The Buneman waves observed in the shock foot have a}
wavevector 
\begin{equation}
k=\frac{c}{\Delta v}\frac{1}{\lambda_{\rm se}},
\label{buneman}
\end{equation}
where $\Delta v$ is the relative speed between incoming electrons and reflected ions, \mpo{and they are excited} when $\Delta v$ exceeds the thermal speed of the electrons, $\Delta v/v_{\rm th}\gg 1$, \jn{\mpo{an expression} that is equivalent to that given in Equation~\ref{unstable}.}
\mpo{The beam of reflected ions is warm, and we expect the electrostatic instability to operate in the kinetic regime and the wave vector of the peak intensity to be aligned with the streaming velocity \citep{Breizman90}.} Figure~\ref{Fourier} displays Fourier \jn{power} spectra of the electric field parallel to the wave vector, $\boldsymbol{e}_k\cdot \boldsymbol{E}$, \jn{calculated in the regions shown in Figure~\ref{Eest}}. The mean field values and its linear gradients were removed from the $E_x$ and $E_y$ maps for this analysis. \jn{Corresponding ion and electron phase-space distributions in $v_x-v_y$ are shown in Figure~\ref{velocities}.} 
\mpo{For run A1 we see a narrow signal at $k_{\parallel}\lambda_{\rm se}\simeq 3.3$ \mpo{in the range $k_{\perp}\lambda_{\rm se}\lesssim 0.7$ (Fig.~\ref{Fourier}a), \jn{that} corresponds to $k \simeq 3.3\,\lambda_{\rm se}^{-1}$ and $\Delta v\simeq 0.3c$}. \jn{Noting that the plasma motion in this case (A) is primarily along the $x$-direction and calculating averaged velocity components we obtain $v_{\rm xi}\simeq-0.15c$ and $v_{\rm xe}\simeq 0.17c$ for reflected ions and incoming electrons, respectively (compare Fig.~\ref{velocities}a). The relative velocity is thus} 
%(upstream electrons start to loose energy because of interaction with reflected ions) 
$\Delta v\simeq 0.31c$}, in good agreement with the value calculated from the Fourier spectrum. \jn{The modes can be clearly identified with the Buneman waves.} 

For the simulation with $\varphi=45^o$ (run B1) \mpo{we see the mode out to $k_{\perp}\lambda_{\rm se}\simeq 4$. The phase-space plot (Fig.~\ref{velocities}b) suggests that the parallel mode should be observed between $\mathbf{k}\lambda_{\rm se}\simeq (3;0)$ and $\mathbf{k}\lambda_{\rm se}\simeq (2.5;3)$, which is again in rough agreement with the range of wave vector for which the Fourier analysis indicates a high intensity of $\boldsymbol{e}_k\cdot \boldsymbol{E}$ waves modes.}

%Unlike \cite{2016ApJ...820...62W} we do not observe a weak oblique mode in the case with $45^o$ magnetic field configuration because of too small $\delta v$ in $y$-direction and also such tiny effects could always be spoiled by non stable character of the shock processes.  \jn{THE ABOVE NEEDS TO BE VERIFIED}
\mpo{For the out-of-plane magnetic field configuration the velocity range of reflected ions inserted in Equation~\ref{buneman} implies strong growth between $\mathbf{k}\lambda_{\rm se}\simeq (1.4;1.4)$ and $\mathbf{k}\lambda_{\rm se}\simeq (0.75;2)$, exactly where we observed it in the Fourier spectra.}

\mpo{The Fourier power spectra also indicate wave intensity at wave vectors not aligned with the streaming velocity of reflected ions. Figure~\ref{Eest}c2 suggests that at least part of that arises from localized reorientation of the wave fronts in $E_x$, probably caused by modulation through large-scale modes.} %These results are in good agreement with the findings of \cite{2012ApJ...755..109M}. 

\mponn{The high electrostatic field amplitudes in run C are somewhat surprising, because there are fewer reflected ions in this run compared to the simulations with $\varphi=0^o$ and $\varphi=45^o$. \citet{2009PhPl...16j2901A} modified an estimate by \citet{1980PhRvL..44.1404I} for the transferrable energy density,
\begin{equation}
\epsilon_0\frac{E^2}{2}\simeq \frac{3\,N_e\,m_e\,\Delta V^2}{8}\,
\left(\frac{m_e}{m_i}\right)^{1/3},
\end{equation}
and we find fairly good agreement with the energy density seen in our simulation ($\sim 70\%$), provided \jnr{that} we use the component of $\Delta V$ that lies in the simulation plane (and is indicated in Fig.~\ref{velocities}). Only for run C (or $\varphi=90^o$) we find the trajectories of reflected ions fully contained in the simulation plane, and so part of the streaming motion in runs A and B cannot excite the Buneman instability, because $\mathbf{k}$ must lie in the simulation plane and $\mathbf{k}\parallel\mathbf{\Delta V}$.}

\mponn{We conclude that the out-of-plane configuration is best suited to fully capture the development of the Buneman waves in a 2D3V simulation, although the adiabatic index is modified in that case. Equation~\ref{trapping} applies in that case, \jnr{and} the velocity difference between the electrons and the reflected ions is 
$\mathbf{\Delta V}¸\simeq 2^{-0.5} \Delta V (1,1,0)$. If the large-scale magnetic field is not strictly out-of-plane, but stands at an angle $\varphi<90^o$ to the simulation plane, then the relative motion is partially rotated out of the simulation plane, and
\begin{equation}
\mathbf{\Delta V}\simeq 2^{-0.5}\Delta V (1,\sin\varphi,\cos\varphi).
\label{deltaV}
\end{equation}
The $z$-component cannot drive Buneman waves, and so we need a larger shock speed to drive the wave intensity in the simulation plane to a level that permits electron trapping. We therefore suggest that formula~\ref{trapping} for the trapping condition be modified to
\begin{equation}
 M_A \geq \sqrt\frac{2}{1+\sin^2\varphi}
 (1+\alpha)  \left(\frac{m_\mathrm{i}}{m_\mathrm{e}}\right)^{\frac{2}{3}} \ .
\label{trappingnew}
\end{equation}
Inserting numbers, including the reflection rates measured in the simulation, we find that in run A we would need $M_A\simeq 39$, and $M_A\simeq 35$ in run B, for efficient trapping. Both simulations are set up with $M_A=31.7$, and so the energy of streaming in the simulation plane \jnr{seems to be} insufficient to drive very strong Buneman modes and permit shock surfing acceleration. Likewise, condition~\ref{trappingnew} would be strongly violated in the $\varphi=0^o$ simulations presented by \citet{2015Sci...347..974M}, and indeed no significant SSA was reported.}

\mponn{Applying a similar modification to the driving condition (Eq.~\ref{unstable}) leads to the expression
\begin{equation}
   M_s \gtrsim \frac{1+\alpha}{\sqrt{2(1+\sin^2\varphi)}} \ \sqrt{\frac{m_i}{m_e}} \ \sqrt{\frac{T_e}{T_i}}\ .
\label{unstablenew}   
\end{equation}
}

\mponn{We also note that the restriction of Buneman waves with $k_z$ component to the $x$-$y$ plane changes the orientation of the potential wells in which electrons can be trapped. Shock surfing acceleration arises from motional electric field that accelerates along the electrostatic equipotential surface of the waves, and a geometrical restriction of $\mathbf{k}$ will also modify the relevant component of the motional electric field, $\mathbf{E}\perp \mathbf{k}$. The dominant effect in our simulations appears to be the reduction in the saturation amplitude though.}

\mpo{The properties of the Buneman instability region in the foot of the \emph{right} shock propagating into the warm plasma are similar to those described above for the small-$\beta_{\rm p}$ \emph{left} shocks. In both cases, the shock reformation imposes cyclic changes to the appearance of the electric-field fluctuations in the foot region, but some general features of Buneman wave fields can be summarized as follows: 
\begin{itemize}
\item The Buneman modes evolve kinetically and are largely, but not perfectly, parallel to the streaming velocity of reflected ions.
\item The streaming direction of reflected ions depends on the orientation of the large-scale magnetic field in 2D simulations, and so the Buneman modes are approximately parallel to the shock normal for the in-plane magnetic field (run A), turn oblique for $\varphi=45^o$ (run B), and are highly oblique for the out-of-plane magnetic-field configuration (run C).
\item In all six runs the wavelengths of Buneman waves are commensurate with the streaming speed of ions, and hence they are smaller for runs A-B in comparison with run C.
\item The region \jnn{with high-intensity coherent waves} is larger and considerably less patchy for the out-of-plane configuration (run C) than it is for an inclined or in-plane magnetic field (runs A and B), irrespective of the plasma $\beta_{\rm p}$.
\mpo{\item In moderate-temperature plasma (runs *2) the peak intensity of Buneman waves is less than that in the cold plasma (runs *1), \jnn{but the surface area occupied by the waves is larger in that case.}}
\end{itemize}
}

\subsection{Electron Acceleration}
The parameters in our simulations \jn{should} provide \jn{suitable} conditions for electron acceleration up to nonthermal energies. \jn{As known from earlier studies of high-Mach-number shocks, the electron injection at such shocks is at least a two-stage process that starts with the SSA in the shock foot and then is followed by additional particle energization processes in the shock ramp and around the overshoot. In this section we describe the electron pre-acceleration at a perpendicular shock. Our two-dimensional numerical experiments with various configurations of the average magnetic field with respect to the simulation plane allow us to observe the injection processes from different perspectives. This in turn enables us to draw conclusions on the nature of electron pre-acceleration and its true efficiency in a fully three-dimensional system.} 

\jn{We start our discussion in Section~\ref{ssa} with a description of the SSA process in the shock foot containing the electrostatic Buneman waves. The analysis is based on results of the (\emph{left}) shocks propagating in cold plasmas with very low plasma beta (runs A1-C1). As noted in Section~\ref{results}, these shocks provide conditions for strongly nonlinear Buneman modes in the shock foot. Subsequent processes of further electron energization at the shock front are described for shocks in plasmas with a
moderate plasma beta (runs A2-C2), at which the injection is more efficient than in the cold plasmas, as we show. The analysis here is presented only for the case of the out-of-plane (run C2, Sec.~\ref{electron-acceleration-runC}) and the in-plane (run A2, Sec.~\ref{electron-acceleration-runA}) magnetic-field configurations, since}
electron acceleration processes \jn{observed in the $\varphi=45^o$ case} are essentially the same as in run A ($\varphi=0^o$).

\subsubsection{Shock-Surfing Acceleration} \label{ssa}

\begin{figure}[htb]
\centering
\includegraphics[width=\linewidth]{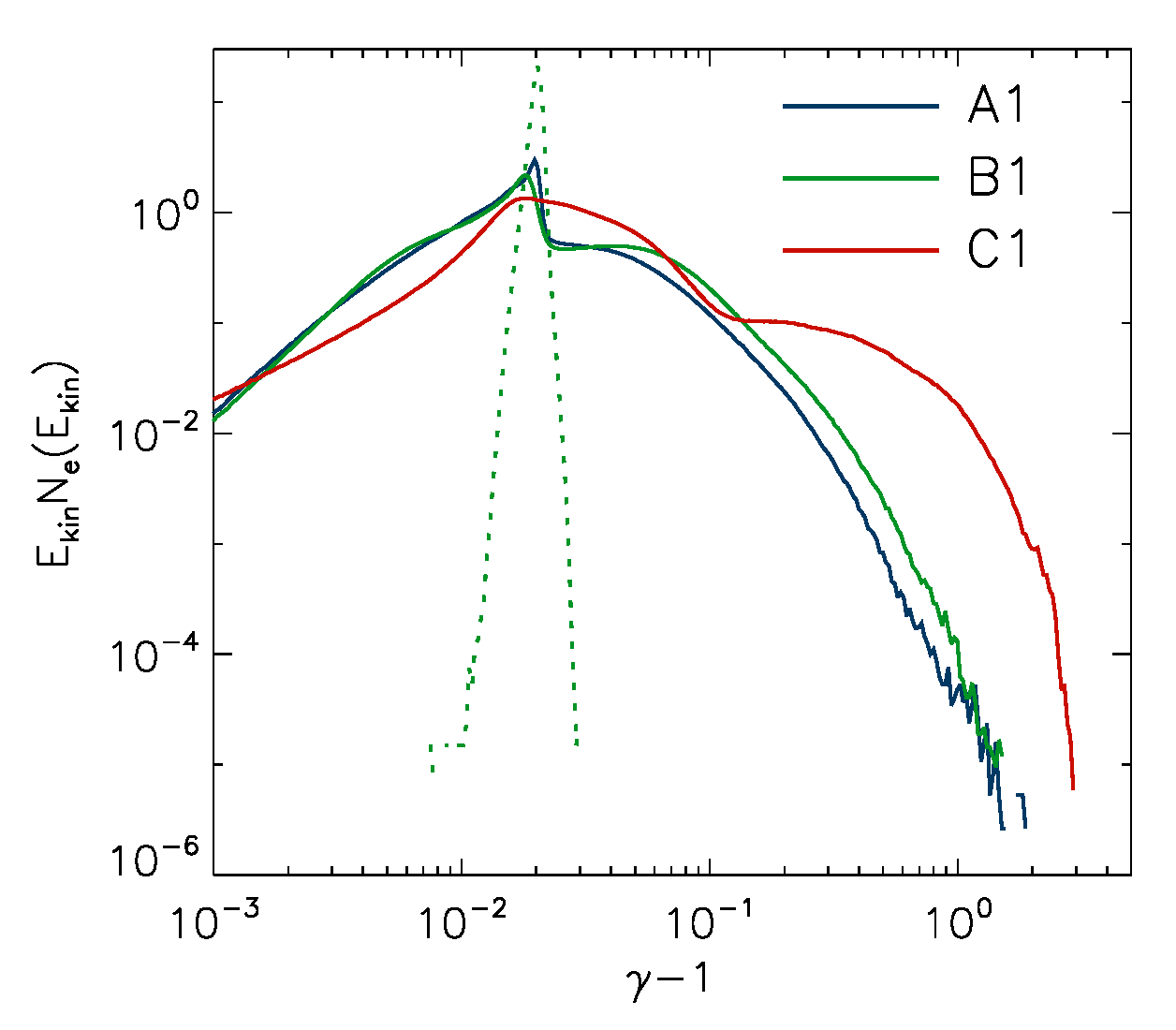}
\caption{\mpo{Kinetic-energy spectra of electrons in the region harboring Buneman waves as presented in Fig.~\ref{Eest} for run A1 (blue), run B1 (green) and run C1 (red). The dotted green line indicates the spectrum of upstream cold plasma electrons (extracted from run B1).}}
\label{bun_acc}
\end{figure}

\begin{figure}[htb]
\centering
\includegraphics[width=\linewidth]{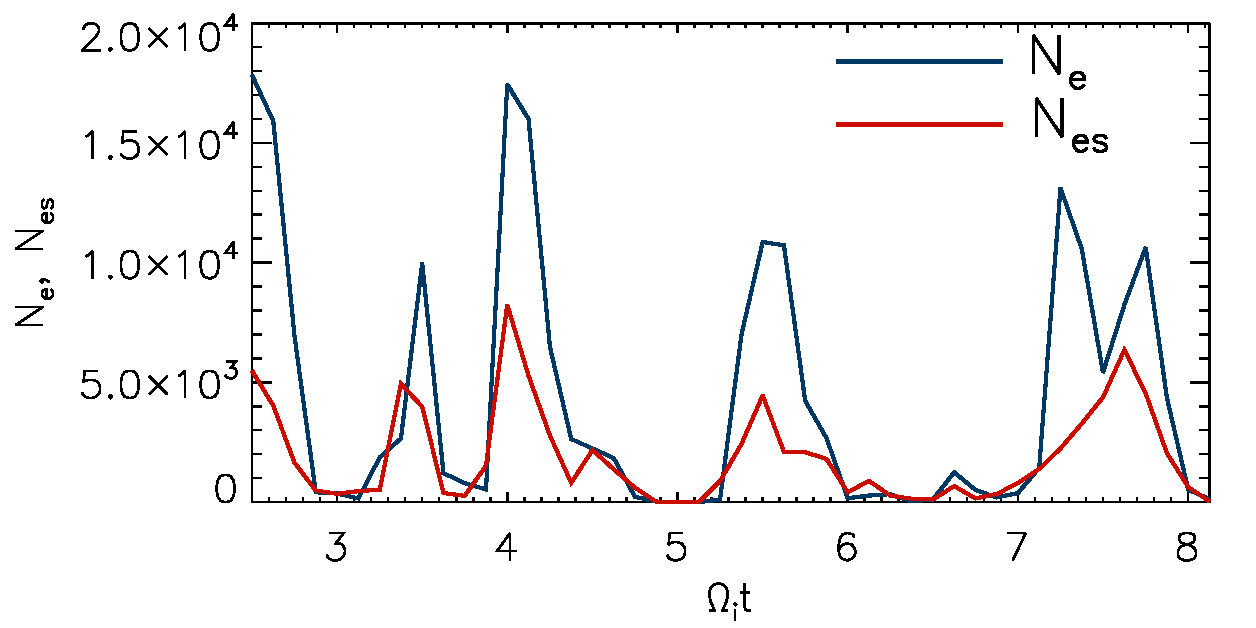}
\caption{\mpo{Temporal profiles of the number of energetic electrons, $N_e(\gamma>1.5)$ (blue line), in comparison to the abundance of strong electrostatic field, $N_\mathrm{es}(E_\mathrm{es} > 0.5\, B_0)$ (red line). {The Spearman rank correlation coefficient is 0.9 with a p-value of $4 \cdot 10^{-17}$.} The figure applies to the region harboring Buneman waves in run A1. }}
\label{bun_ele_esp}
\end{figure}

Figure~\ref{bun_acc} \mpo{shows kinetic-energy spectra of electrons residing in the shock-foot regions depicted in Figure~\ref{Eest}. For all magnetic-field configurations supra-thermal spectral tails are produced by SSA. However,  the} high-energy \jn{portion of the electron spectrum} for run C1 contains a significantly larger number of particles than \jnn{that} found for runs A1 and B1. \jn{Specifically,} there are about 2200, 7600 and $3.4\times 10^6$ electrons with $\gamma>1.5$ for runs A1, B1, and C1, respectively.  \jn{These spectral differences can be explained by the amplitude and filling factor of the Buneman waves in the shock foot, as discussed in Section~\ref{Buneman}. In runs A1 and B1, the wave zone is narrow, and the \jnn{wave} amplitude barely \mpon{sufficient to} trap relativistic electrons.}
%Low intensity and small region with electrostatic waves explains the small number of energetic electrons for runs A1 and B1 in comparison  with run C1. Slightly better electron energization in run B1 in comparison with run A1 can be explained by larger intensity of electrostatic waves. 
\jn{On the other hand,} in run C1, the Buneman waves occupy an almost three times larger region and \jn{are more intense and coherent}. We find that \jn{many} electrons \jn{approaching} the foot \jn{from the far upstream} can interact with electrostatic waves twice or even three times \jn{before being advected toward the shock. This multiple SSA processes enable the particles} to achieve energies up to $\gamma\sim 4$ \citep{2009ApJ...690..244A,2012ApJ...755..109M}, \jnn{before they are further energized in the shock ramp}.

\mpo{There is clear correlation between the number of pre-accelerated electrons and the occurrence of intense electrostatic field. Figure~\ref{bun_ele_esp} displays for run A1 the temporal development of the number of electrons with Lorentz factor $\gamma>1.5$ and the number of computational cells with electrostatic field strength commensurate with or stronger than the initial homogeneous magnetic field. A~Spearman rank test yields correlation coefficients in the range $0.7$--$0.9$ for all 6 simulation runs with very small p-values, indicating a good correlation.}

\subsubsection{Electron acceleration (run C2)} \label{electron-acceleration-runC}

The main stages of electron acceleration for the out-of-plane magnetic-field configuration \jn{(run C2)} are shown in Figure~\ref{trac_dens_run_C},  \jn{in which we trace the trajectory and the energy evolution of a typical particle that becomes energized at the shock.}
At time $t_a$ the electron resides in the Buneman-wave region and \mpo{is subjected to the SSA process, through which its} energy reaches $\gamma \simeq 2.5$.  \jn{In this example, the electron does not experience another SSA cycle but}
starts to gyrate \jn{around the mean magnetic field} and \mpo{passes through the shock ramp} towards the overshoot. \mpo{The projection of its gyromotion on the convective electric field in the shock ramp imposes quasi-sinusoidal variations in} its kinetic energy. \mpo{A significant energy increase arises only beyond time $t_b$ when the electron approaches the overshoot. The highest energy \jnn{of $\gamma \simeq 6$} is reached around the time of highest compression in the overshoot, shortly after $t_c$, and upon entering %Next, the particle starts to interact with the shock overshoot. Because of the strong magnetic field compression the}  electron \jn{quickly} gyrates in the overshoot and undergoes the gradient-$B$ drift acceleration (time $t_3$ in Fig.~\ref{trac_dens_run_C}c). Its kinetic energy rapidly grows up to the value of $(\gamma-1) \simeq 5$, \jn{achieved at time in which} the electron \jn{still resides in a} region with the maximum magnetic field \jn{strength}. \jn{Upon crossing the shock front toward 
the downstream region, in which the plasma compression is much smaller than in the overshoot, the energy of the electron slowly decreases. 
The acceleration beyond the SSA phase is thus largely adiabatic. In fact, after time $t=t_a$, 
the average value of the magnetic moment of the electron remains constant \citep[compare][]{2009ApJ...690..244A}.}

To examine \jn{the dynamics} of the acceleration processes, we follow the temporal evolution of the spectrum for a selected portion of upstream electrons \jn{traversing the shock structure}.\mpo{
Figure~\ref{ele-acc-run_C}a displays electron spectra at four points in time, and the panels b-e indicate the location of the particles for each of the four instances.} Each spectrum is fitted with a relativistic Maxwellian (dashed lines in Fig.~\ref{ele-acc-run_C}a), and the fraction of non-thermal electrons is estimated. \mpo{A supra-thermal tail is evident in the spectra already after passage though the Buneman wave field at the shock foot (Fig.~\ref{ele-acc-run_C}b). During their further transport through the shock the electrons are accelerated to very high energies, and at the same time their bulk temperature increases.} The nonthermal electron fraction is about $4.8\pm 0.2\%$, carrying $\sim26\%$ of the total electron energy. \mpo{It remains roughly constant once the particles pass through the overshoot and propagate toward the downstream region, where the bulk temperature decreases and the spectral tails become less prominent than those at the overshoot. This behavior provides general support for the notion that the acceleration beyond the SSA phase is adiabatic, that we demonstrated for a single particle above. Particle heating is achieved through bulk plasma compression in the shock that is strongest at the overshoot.} %Any departures from a strictly adiabatic character of the spectral evolution after the SSA phase result mainly from the fact that the electrons selected for the spectra calculations at phases shown in Figures~\ref{ele-acc-run_C}c-e occupy extended regions of the shock transition with varying plasma compression. Note that additional processes of the betatron acceleration, induced by the shock-reformation, can also contribute to the final particle spectra \citep{2012ApJ...755..109M}.} 

\begin{figure}[htb]
\centering
\includegraphics[width=0.98\linewidth]{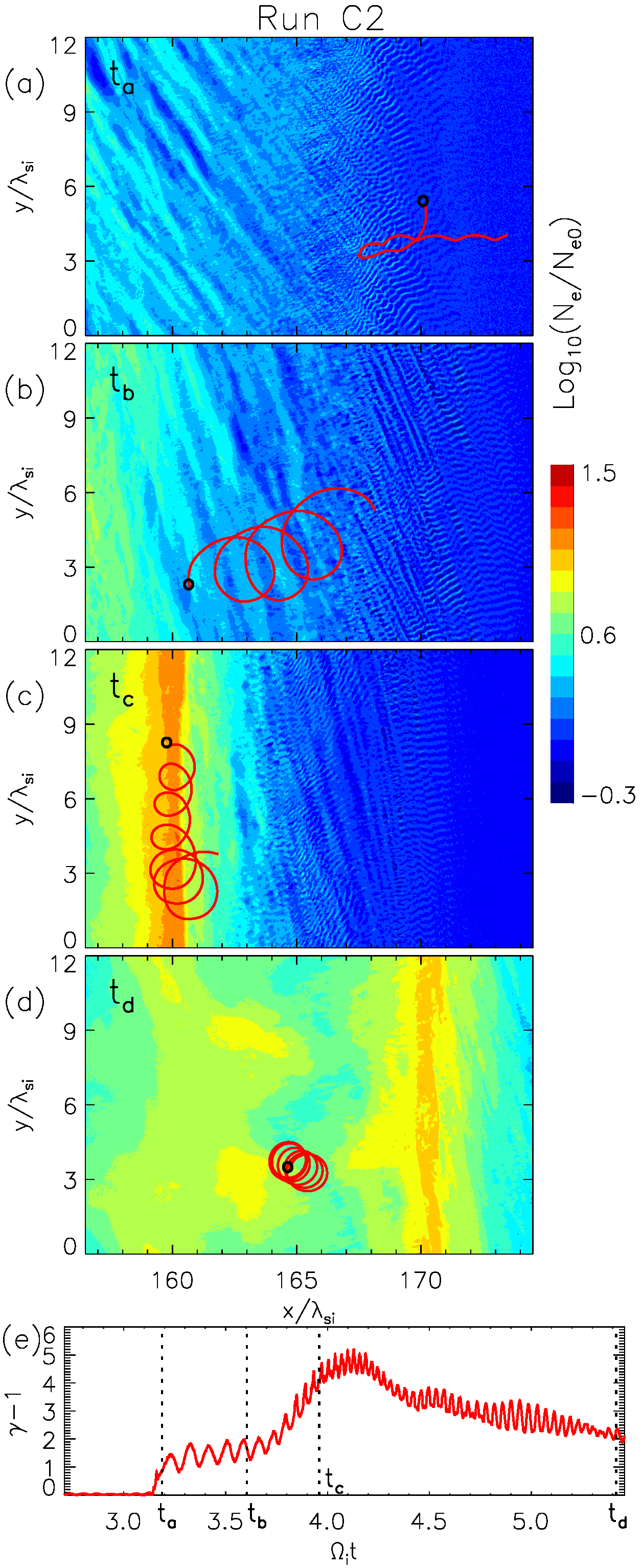}
\caption{\mpo{Trajectory segments of a high-energy electron from run C2 superposed on the electron density map at four time intervals. The panels (a),(b),(c),and (d) display the state of the system at the times $t_a$ (a), $t_b$ (b), $t_c$ (c) and $t_d$ (d) that are marked in panel (e), and the black circles indicate the position of the electron at these moments. The red lines give their trajectory history for the time span  $0.3\Omega_i^{-1}$. The color scale for the normalized electron density is logarithmic. In panel (e) we present the temporal development of the kinetic energy of the electron.} }
\label{trac_dens_run_C}
\end{figure}

\begin{figure}[htb]
\centering
\includegraphics[width=0.99\linewidth]{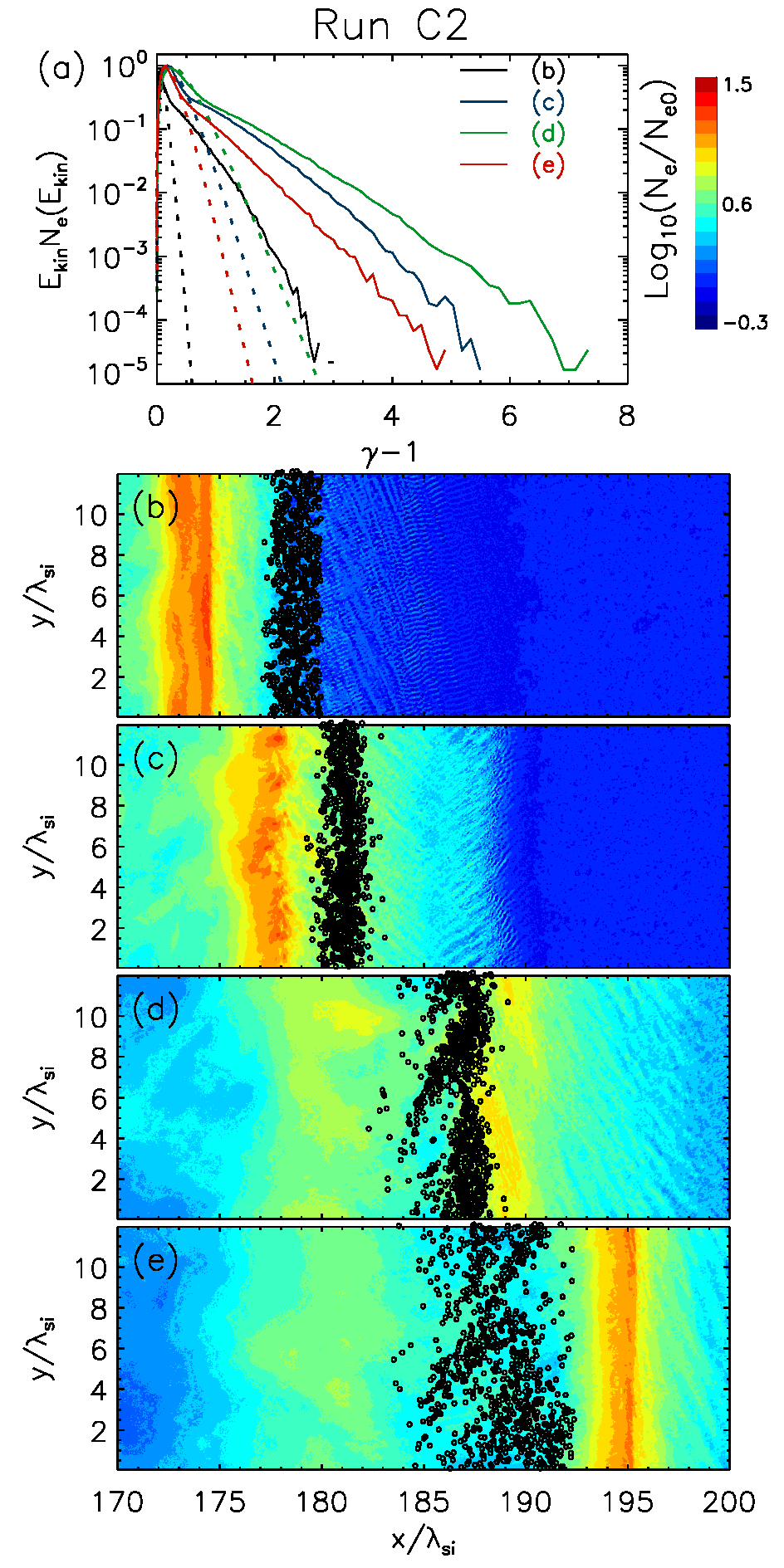}
\caption{\mpo{Panel (a): Spectral evolution of electrons extracted from the downstream region for run C2 compared to fits of a relativistic Maxwellian plotted as dotted lines. Panels (b), (c), (d) and (e) present density maps of the shock region together with the positions of electrons selected in the upstream region. The panels display the status at time $t=5.625\Omega^{-1}$, $6.25\Omega^{-1}$, $7\Omega^{-1}$, $7.375\Omega^{-1}$, respectively. The logarithmic color legend for the density maps is placed near panel (a).}}
\label{ele-acc-run_C}
\end{figure}

\subsubsection{Electron acceleration (run A2)} \label{electron-acceleration-runA}

%\begin{figure}[htb]
%\centering
%\includegraphics[width=8.5cm]{trac_dens_run_A3.eps}
%\caption{Trajectory of high energy electron from run A superposed on the electron density map (panels (a),(b),(c)). Snapshots are taken at time moments $t_1$ (a), $t_2$ (b) and $t_3$ (c). Color presents normalized electron density in logarithmic scale, black circle - position of electron for the moments $t_1$ (a), $t_2$ (b) and $t_3$ (c), red lines - trajectory history for $0.2\Omega_i^{-1}$. $(\gamma-1)$ history of electron is presented in panel (d). Dashed lines designate moments $t_1$, $t_2$ and $t_3$.}
%\label{trac_he}
%\end{figure}

\begin{figure}[htb]
\centering
\includegraphics[width=0.98\linewidth]{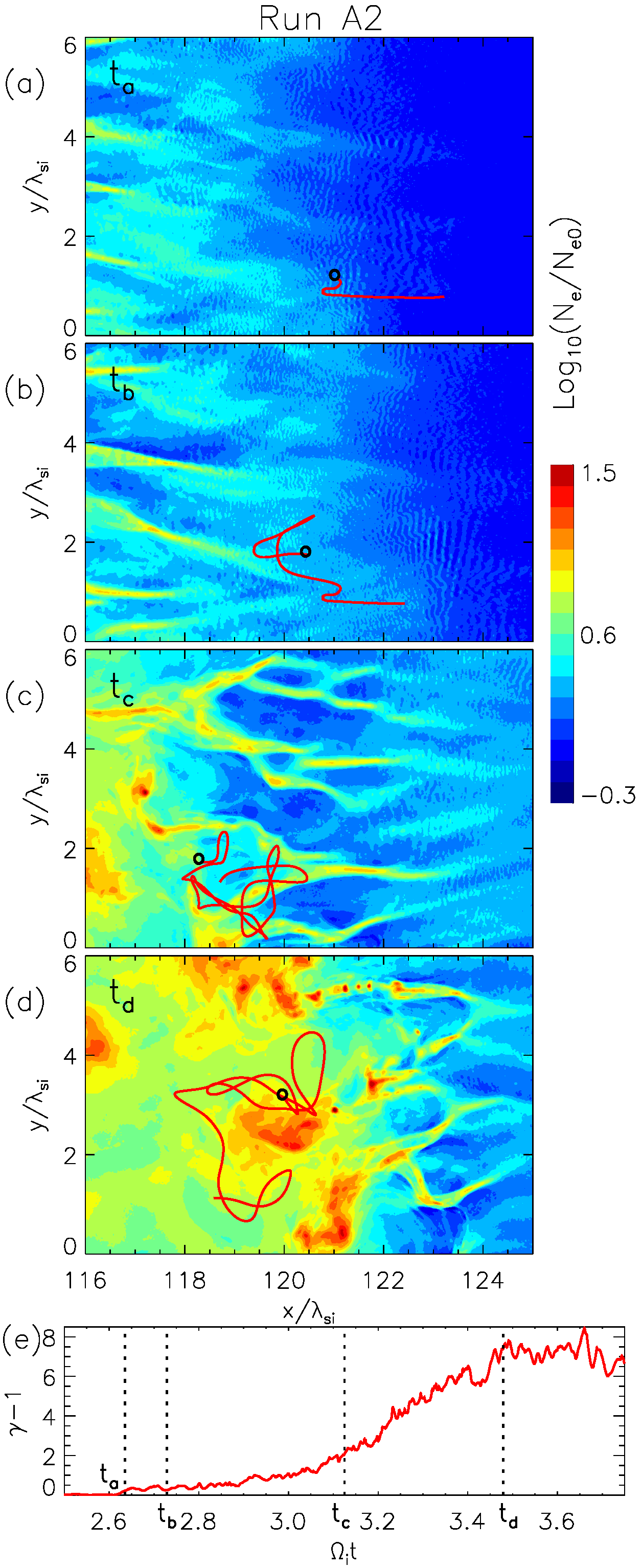}
\caption{Trajectory segments of a high-energy electron from run \jnn{A2} superposed on an electron-density map at four time intervals. \mpo{In panel (e) we present the temporal development of the kinetic energy of the electron and mark four points in time. The panels (a),(b),(c),and (d) display the state of the system at these times} ($t_a$ (a), $t_b$ (b), $t_c$ (c) and $t_d$ (d)), and the black circles indicate the position of the electron at these moments. The red lines give their trajectory history for the time span  $0.2\Omega_i^{-1}$. The color scale for the normalized electron density is logarithmic. }
\label{trac_he2}
\end{figure}

\begin{figure}[htb]
\centering
\includegraphics[width=\linewidth]{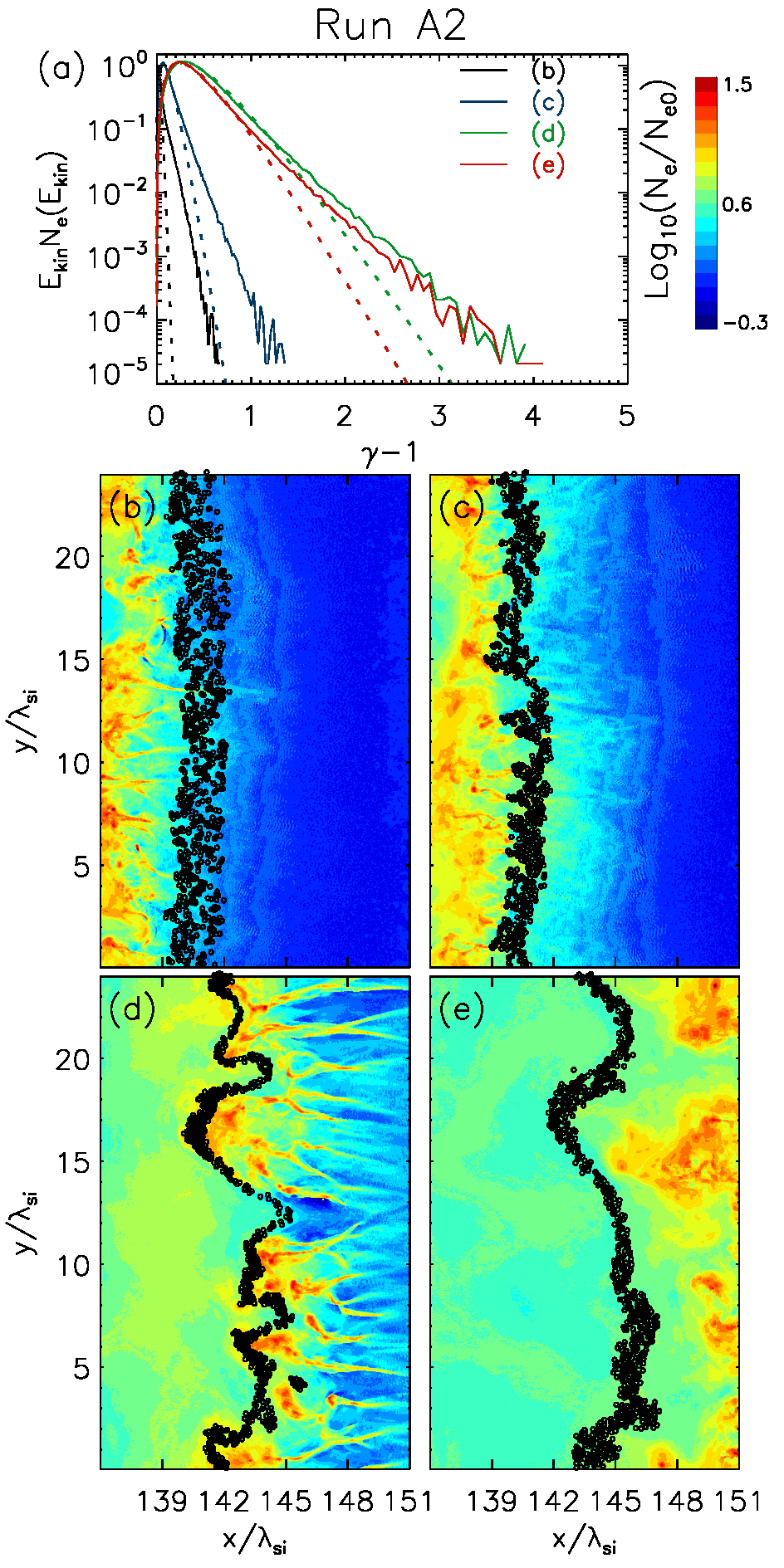}
\caption{\mpo{The top panel (a) displays the spectral evolution of electrons selected in the downstream region for run A2. The dotted lines indicate adaptions of a relativistic Maxwellian. Panels (b), (c), (d) and (e) present density maps of the shock region and the positions of the selected electrons at the time $t=5.75\Omega^{-1}$, $5.875\Omega^{-1}$, $6.625\Omega^{-1}$, $7.375\Omega^{-1}$. Density is presented in logarithmic scale, and the color legend is placed near panel (a).}}
\label{ele-acc-run_A}
\end{figure}

\mpo{Electron acceleration for the case with the in-plane magnetic-field configuration (run A2) has been analysed in a similar way as that for out-of-plane field (Sec.~\ref{electron-acceleration-runC})}. Figure~\ref{trac_he2} follows the trajectory and the kinetic-energy history of \jn{a typical particle acquiring high energy in interactions within the shock structure, and Figure~\ref{ele-acc-run_A} displays spectra of a selected electron population traversing the shock.
% (see Fig.~\ref{ele-acc-run_C}). 
\mpo{To be noted from the figures is that for the in-plane magnetic field} the SSA process in the shock foot results in moderate particle energization. The Lorentz factor of the electron increases only to $\gamma \simeq 1.5$ while it resides in the Buneman zone at the shock foot (time $t_a$ in Fig.~\ref{trac_he2}a), and the supra-thermal tail in the spectrum in Figure~\ref{ele-acc-run_A}a does not reach $\gamma= 2$. Subsequent acceleration (Figs.~\ref{trac_he2}b-d) in the shock structure is markedly different from that observed in the case with the out-of-plane magnetic field: the particle interacts many times with moving magnetic structures, essentially undergoing a stochastic (second-order Fermi) acceleration processes. Its energy increases steadily, and the rate of the energy gain is larger in the shock overshoot region (between times $t_c$ and $t_d$ in Fig.~\ref{trac_he2}e) than in the shock ramp
(time range approximately from $t_b$ to $t_c$ in Fig.~\ref{trac_he2}e), \mpo{because the turbulent magnetic field is stronger at the overshoot. In the end, the sample electron reaches a maximum energy of about $\gamma \simeq 9$ and retains that energy while it is advected into the downstream region of the shock. Electron injection for the in-plane and the $\varphi=45^o$ field configuration thus mainly} involves irreversible non-adiabatic acceleration processes.}

%the we examined trajectories of electrons with the highest final energies. In figure~\ref{trac_he} trajectory and the kinetic energy evolution for the high energy electron are presented. 
%Snapshot~\ref{trac_he}(a) demonstrates the first stage of electron energization by Buneman instability, the electron gains about $0.5mc^2$ that is enough to be a nonthermal particle and then (fig.~\ref{trac_he}(b)) it has a possibility to be accelerated via second-order Fermi processes interacting with magnetic filaments and structures. When the electron goes to the overshoot region magnetic field strength caused by beam-Weibel instability is becoming larger and acceleration through interaction with turbulent magnetic structures becomes more efficient (fig.~\ref{trac_he}(c)). The selected electron reaches maximal energy about $(\gamma-1) \simeq 8$ in the overshoot region. 
%Trajectory analysis of branch of high energy electrons shows that dominant acceleration process is Fermi like chaotic interaction with magnetic shock structures.

\mpo{The observation of non-adiabatic acceleration is supported by an analysis of the electron spectra in Figure~\ref{ele-acc-run_A} that remain largely unchanged once particles reach their maximum energies and are transmitted downstream (Figs.~\ref{ele-acc-run_A}d-e). Small differences in these spectra may result in part from the plasma decompression behind the overshoot, but most probably they reflect the shock reformation. Note, that scattering of energetic particles is accompanied with significant bulk particle heating. As a result, the fraction of nonthermal electrons in the downstream spectrum is about an order of magnitude less than that obtained in the $\varphi=90^o$ case. The spectra also decay at smaller energies, compared with the case of run~C2.}

\mpo{There is some uncertainty in the fraction of nonthermal electrons, because during the shock transition the particles disperse, and so electrons initially confined in a narrow range of $x$ coordinates are distributed over an $x$-range of $10\ \lambda_\mathrm{si}$ once they are in the downstream region, in particular for run C2 ($\varphi=90^o$). The nonthermal fractions are calculated using the Maxwellian fits to the low-energy spectra that are presented in Figures~\ref{ele-acc-run_C}a and~\ref{ele-acc-run_A}a. We find 8.6\%, 6.4\%, 5.9\%, 4.8\% for the snapshots (b), (c), (d), (e) in Figure~\ref{ele-acc-run_C}a, i.e., the out-of-plane configuration, whereas for run A and an in-plane magnetic field we obtain 0.67\%, 0.21\%, 0.31\%, 0.41\% (for snapshots (b), (c), (d), (e) in Figure~\ref{ele-acc-run_A}a). The increase in the nonthermal fraction between the shock ramp, the overshoot, and the downstream region of the simulation with the in-plane configuration (run A) reflects the non-adiabatic acceleration processes that appear to operate near the overshoot. In contrast, very efficient acceleration by SSA is observed in the shock foot for out-of-plane magnetic field, and in the shock ramp and at the overshoot we loose nonthermal energy by randomization and heating.} 

\mpo{The shocks in the B runs (for $\varphi=45^o$) essentially behave like those in the simulations with the in-plane magnetic field (A runs), and so the contents of this subsection also applies to the shocks in B runs.}

%Figure~\ref{ele-acc-run_A} displays spectra of the selected electrons for the different time steps and positions of electrons for these time steps. The first stage of acceleration is acceleration by electrostatic Buneman mode in the foot region (panel (b)) and formation of nonthermal tail is almost finished after crossing Buneman instability region. Then we observe mainly heating of plasma with constant nonthermal fraction. Actually if the main acceleration process is Fermi process then we do not expect growing of nonthermal electron fraction. Between snapshots (d) and (e) we see small spectra relaxation. Finally nonthermal fraction equals $(0.4\pm0.05)\%$ and these electrons hold about $2.4\%$ of total energy. 

\subsubsection{The influence of shock reformation} \label{reformation-nonthermal}

\mpo{In section~\ref{reform} we discussed the \jnn{significant} influence of the cyclic self-reformation on the structure and speed of the shock, the intensity of Buneman waves, and subsequently on particle acceleration. \jnn{Another} consequence is that the downstream particle distributions \jnn{are} not uniform, and the choice of region matters from which we extract particle spectra.}

\begin{figure}[htb]
\centering
\includegraphics[width=\linewidth]{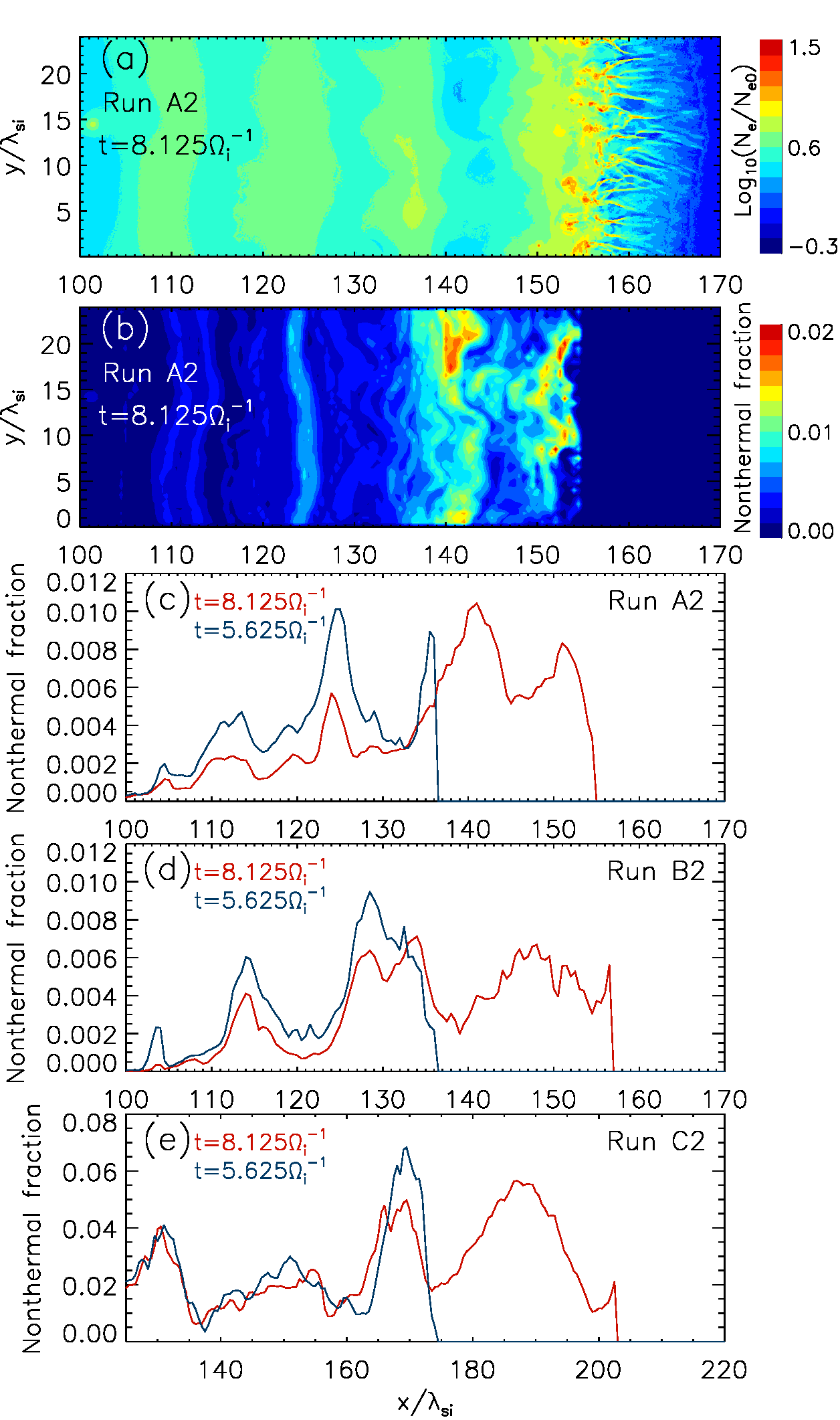}
\caption{Normalized electron density in logarithmic scale (a) and nonthermal electron fraction (b) for run A2 at time $t=8.125\Omega_i^{-1}$. \mpo{The three lower panels display profiles of nonthermal electron fraction at time $t=5.625\Omega_i^{-1}$ (blue) and at $t=8.125\Omega_i^{-1}$ (red)} for run A2 (c), run B2 (d) and run C2 (e).}
\label{spectra-map}
\end{figure}

\mpo{In the downstream region, even in small regions of size $0.5\lambda_{si}\times 0.5\lambda_{si}$ \jnn{we find} a large number of \jnn{computational} particles (\jnn{on average} $2\times 10^5$ for runs A2/B2 and $1.5\times 10^5$ for run C2). \jnn{Hence, \emph{local}} electron spectra can be calculated with reasonable precision. Likewise, we can evaluate the \jnn{\emph{local}} temperature by fitting a relativistic Maxwellian and determine the fraction of nonthermal electrons.
Figure~\ref{spectra-map} shows maps of the electron density and nonthermal electron fraction in the downstream region of the \jnn{moderate}-$\beta_{\rm p}$ shock in run A2. 
To be noted from the figure is the inhomogeneous structure of the downstream region that results from shock reformation. Regions of higher density are signatures of the reformation \jnn{phases} at which the shock overshoot had the highest density. The nonthermal electron fraction is also nonuniform (Fig.~\ref{spectra-map}b). The simulation is too short to permit homogenization of the downstream region, implying that it will be achieved only very far downstream.}

\mpo{Also displayed in Figure~\ref{spectra-map} are profiles of the nonthermal electron fraction for all \jnn{moderate}-$\beta_{\rm p}$ runs. \mponn{To be noted from the figure are the large variations in the nonthermal fraction in the downstream region. They reflect variations in both the bulk temperature and the number of high-energy electron that do not coincide.}
In panel (c) we present profiles (averaged over $y$\jnn{-direction}) of the nonthermal electron fraction in run A2 at time $t=5.625\Omega_i^{-1}$ (blue) and at $t=8.125\Omega_i^{-1}$ (red). One can see four maxima in the nonthermal fraction (around $x/\lambda_{si}\simeq$ 112, 124, 141, and 152) for the red line that trace back to passage through the most intense electrostatic-wave field in the foot region. The blue line displays the nonthermal fraction at an earlier time ($t=5.625\Omega_i$), and we note that the nonthermal fraction at a fixed location was higher at earlier times, indicating a decrease with time of the abundance of pre-accelerated electrons in the downstream region. We observe the same trend for run B, which may explain the marginal nonthermal population found in the far-downstream region in the simulation of \citet{2016ApJ...820...62W}. It is remarkable that we do not observe a similar loss of nonthermal electrons in the simulation with the out-of-plane magnetic field (Fig.~\ref{spectra-map}e).} \mponn{There the amplitude of variations in temperature and electrons density is also larger and somewhat correlated, which suggests that homogenization is not as efficient as for runs A and B.}

\subsubsection{Spectra in the downstream region} \label{spectra-down}

\mpo{We conclude the presentation of \jnn{our} results with the energy spectra of electrons in the downstream region. For that purpose we chose a region behind the overshoot that contains particles processed over two cycles of the shock reformation. The extent of \jnn{this} region is $ 2\times 1.55\Omega_i^{-1} \times v_{\rm sh}\simeq25 \lambda_{\rm si}$ for runs A/B and $37.5\lambda_{\rm si}$ for run C (recall that the shock speed is higher in run C). }

%\begin{figure*}[!t]
%\centering
%\includegraphics[width=5.8cm]{spectra_downstream-A4.eps}
%\includegraphics[width=5.8cm]{spectra_downstream-B4.eps}
%\includegraphics[width=5.8cm]{spectra_downstream-C4.eps}
%\caption{\ab{PLOTS FOR DISCUSSION. Spectra for cold and warm plasma slabs. Run A.}}
%\label{down-sp-runs}
%\end{figure*}
\mpo{Electron spectra for the in-plane ($\varphi=0^o$), $\varphi=45^o$, and the out-of-plane ($\varphi=90^o$) configuration of the magnetic field and $\beta_{\rm p}=0.5$ are presented in Figure~\ref{spectra-0-45-90}a.
In all simulations we observe electrons with \jnn{Lorentz factors} up to $\gamma\approx 9$. The main difference between the spectra is at low energies, \jnn{at which} we can fit relativistic Maxwellians to represent the bulk of electrons, shown here \jnn{with} dashed lines. To be noted is the variation in the plasma temperature that results from the choice of the magnetic-field configuration. In the lower panel (b) of Figure~\ref{spectra-0-45-90} we display spectra in energy scaled to the plasma temperature. Whereas for $\varphi=0^o$ and $45^o$ we find almost indistinguishable spectra in rescaled energy, those for run C with the out-of-plane field feature a much more pronounced \jnn{spectral} tail.}

\mpo{Table~\ref{table-spectra} summarizes our findings: the downstream temperature, the nonthermal electron fraction, and the number of energetic electrons are presented for all runs. Note, that besides the fit uncertainty in the temperature there is a spatial variation of plasma temperature in the downstream region, and so we consider the plasma temperature for low and \jnn{moderate} $\beta_{\rm p}$ the same within the uncertainties. As was shown in Sections~\ref{ssa}, \ref{electron-acceleration-runC} and \ref{electron-acceleration-runA}, the number of pre-accelerated electrons and the final abundance of nonthermal electrons depends on the efficiency of acceleration by the Buneman waves. The electron temperatures are higher by a factor 2--4 than those predicted by the Rankine-Hugoniot jump conditions, indicating that significant bulk heating has occurred \jnn{\citep[compare][]{2012ApJ...755..109M}}.}

\mpo{It is remarkable that there is no direct relation between large-amplitude electrostatic field and the fraction of nonthermal electrons. While the uncertainties in the determination are sizable, we find a higher nonthermal fraction for \jnn{moderate}-$\beta_{\rm p}$ shocks, whatever the orientation of the large-scale magnetic field.}

\mpo{We do observe a correlation between the abundance of strong electrostatic field and the presence of a high-energy tail in the final downstream spectrum, expressed as either the maximum energy or the number of high-energy electrons with $\gamma>3$.
For the number of high-energy electrons one can find the same trend in all simulations: if the number of grid points with high-amplitude electrostatic field, $N_\mathrm{es}(E_\mathrm{es} > 0.5\, B_0)$, is high, then the number of high-energy electrons, $N_e(\gamma>3)$, is also high. Comparing shocks propagating in the cold and the warm plasma (low and \jnn{moderate $\beta_{\rm p}$, respectively}), we find for run A $N_\mathrm{es, cold}(E_\mathrm{es} > 0.5\, B_0)>N_\mathrm{es, warm}(E_\mathrm{es} > 0.5\, B_0)$, and indeed we observe a higher abundance of high-energy electrons for the low-$\beta_{\rm p}$ shock. For run B, both the abundance of intense electrostatic field and the spectral tails are similar for low and \jnn{moderate} $\beta_{\rm p}$. For run C, the correlation holds, but now we find strong electric field more rarely at shocks propagating into the cold plasma, and there are fewer energetic particles there than at the \jnn{moderate}-$\beta_{\rm p}$ shock. We can conclude that the population size of high-energy electrons (but not necessarily the non-thermal fraction of electrons) is determined by energization in the Buneman zone.
%The temporal evolution of selected particles during their crossing of shock region appears to be adiabatic (here we understand that acceleration processes in shock region are adiabatic only in run C, while in runs A and B are not adiabatic HOW CAN IT LOOK LIKE ADIABATIC IF IT ISN'T???).
} 

%Relativistic Maxwellian fit gives the electron temperatures of $kT/mc$ and they are presented in table~\ref{table-spectra}. Temperature differences between cold and warm plasma cases have stochastic character and caused mainly by variations of plasma temperature in downstream region. Smaller downstream temperature in run C could be explained by smaller compassion factor of the downstream plasma.

%   \begin{table}
%      \caption[]{Nonthermal electron fraction in downstream spectra}
%         \label{table-spectra}
%     $$ 
%\begin{array}{p{0.1\linewidth}llll}
%%\hline
%\noalign{\smallskip}
%$Run$   & \varphi &  $cold shock$ & $hot shock$  \\
%\noalign{\smallskip}
%\hline
%\noalign{\smallskip}
%A  & 0^o  &  0.002\pm0.001  &  0.007\pm0.001  \\
%B  & 45^o &  0.002\pm0.001  &  0.005\pm0.001  \\
%C  & 90^o &  0.04\pm0.01    &  0.07\pm0.01   \\
%\noalign{\smallskip}
%\hline
%\end{array}
%     $$ 
%   \end{table}

   \begin{table}
      \caption[]{Downstream spectra parameters.}
         \label{table-spectra}
\centering
\begin{tabular}{lccccl}
\hline
\hline
\noalign{\smallskip}
Run   & $\varphi$ & $\beta_\mathrm{p}$ & NTEF (\%) & $N_e(\gamma>3) (\%)$ & $kT/mc^2$  \\
\noalign{\smallskip}
\hline
\noalign{\smallskip}
A1  & $0^o$   & 0.0005 & $0.2\pm0.1$  &  0.1  & 0.053  \\
A2  & $0^o$   & 0.5 & $0.7\pm0.1$  &  0.06  & 0.043    \\
\noalign{\smallskip}
\hline
\noalign{\smallskip}
B1  & $45^o$  & 0.0005 & $0.2\pm0.1$  &  0.11  & 0.054  \\
B2  & $45^o$  & 0.5 & $0.5\pm0.1$  &  0.1  & 0.049    \\
\noalign{\smallskip}
\hline
\noalign{\smallskip}
C1  & $90^o$  & 0.0005 & $4\pm1$  &  0.12  & 0.032  \\
C2  & $90^o$  & 0.5 & $7\pm1$  &  0.3  & 0.03    \\
\noalign{\smallskip}
\hline
\end{tabular}
\smallskip
\tablecomments{\mpo{Comparison of the characteristics of the energy distribution of electrons in the downstream region of all six simulated shocks. NTEF denotes the nonthermal electron fraction.}} 
   \end{table}

\begin{figure}[htb]
\centering
\includegraphics[width=\linewidth]{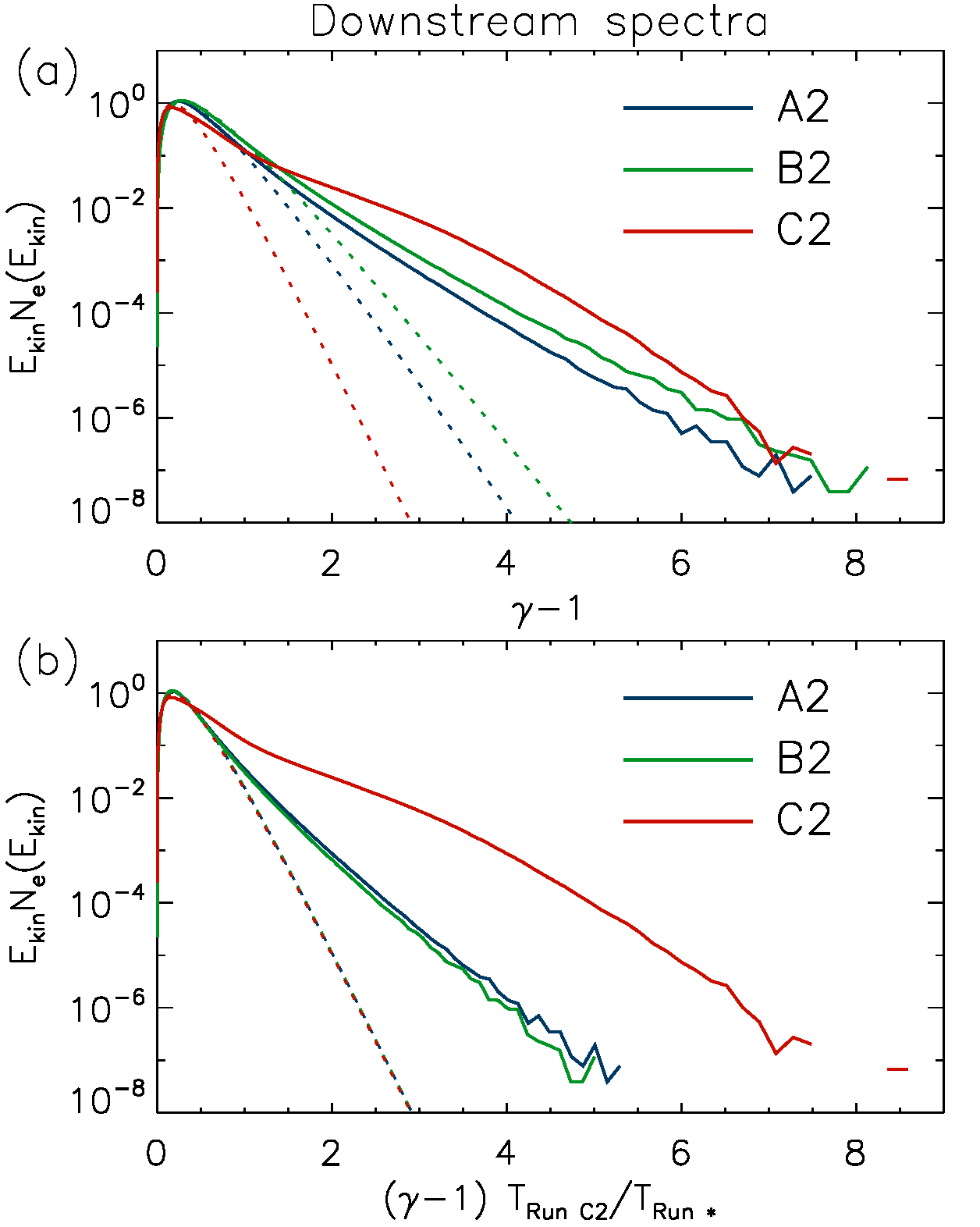}
\caption{\mpo{In the top panel (a) we show electron spectra in the downstream region of the \jnn{moderate}-$\beta_{\rm p}$ shocks (a). The dashed lines represent fits of a \jnn{relativistic} Maxwellian to the low-energy spectra. The bottom panel (b) displays the distribution in rescaled energy in units of the downstream temperature. Blue lines correspond to run A2, green curves to run B2, and red curves to run C2.}}
\label{spectra-0-45-90}
\end{figure}

\mpo{Efficient electron acceleration during passage through the shock ramp and overshoot will weaken the relation between the nonthermal fraction of electrons and the efficiency of SSA. In the $\varphi=90^o$ case with the out-of-plane magnetic field the electron transport beyond the Buneman zone is adiabatic. 
\mpon{Consequently, the distribution of electrons in magnetic moment (instead of their energy) is constant in time,} 
%and consequently the distribution of particles in magnetic moment is constant in time, 
as is the numerical relation between thermal and nonthermal electrons. Then the number of nonthermal electrons should be a direct tracer of the SSA efficiency. In $\varphi=0^o$ and $\varphi=45^o$ configurations \jnn{second-order} Fermi-like processes operating in the shock ramp can change the number ratio of the thermal bulk and the nonthermal population. Further studies are needed to explore the nature of non-adiabatic acceleration processes, for example \jnn{spontaneous turbulent magnetic} reconnection \citep{2015Sci...347..974M}, for which the shock reformation provides an additional complication. We defer this issue to a further publication.}

%\begin{figure}[htb]
%\centering
%\includegraphics[width=7cm]{mag_mom_R_sel_part2_run2a.png}
%\caption{Magnetic moment and gamma-1 spectra for electrons (Run 2a)}
%\label{mag-mom-gamma-spectra-run_2a}
%\end{figure}

\section{Summary and discussion} \label{summary}

%\ab{Under construction...}
\mpo{In this paper we present results of our 2D3V PIC simulations of electron injection and acceleration processes in non-relativistic perpendicular collisionless shocks with high Alfv\'enic Mach number. 
%
%We explored the efficiency of electron energization for three different orientations of the large-scale magnetic field \jnn{and conditions of the cold and moderate-temperature plasmas.}}
%
%The simulation parameters are chosen to \mpo{permit the growth of the Buneman waves and the trapping of electrons at the shock foot \citep[see][]{2012ApJ...755..109M}. We investigate shocks in low and \jnn{moderate}-$\beta_{\rm p}$ plasma for three angles, $\varphi$, between the large-scale magnetic field and the simulation plane. 
%
The simulation parameters are chosen to \mpo{permit the growth of the Buneman waves and the trapping of electrons at the shock foot \citep[see][]{2012ApJ...755..109M}. \jnn{We explored the efficiency of electron energization for shocks in} plasmas with low and \jnn{moderate} $\beta_{\rm p}$ \jnn{and} for three \jnn{orientation} angles, $\varphi$, between the large-scale magnetic field and the simulation plane. The Alfv\'enic Mach number is $M_A=31.8$ for simulations with angle $\varphi=0^o$ and $45^o$, whereas for $\varphi=90^o$ we have $M_A=35.6$ on account of a larger effective adiabatic index. We follow the temporal evolution of shock structures for eight ion gyrotimes, $t=8 \Omega_i^{-1}$,} to study the influence of the shock self-reformation on particle acceleration.

Our results can be summarized as follows:
\begin{itemize}

\item The collision of two plasma slabs leads to the formation of a \jnn{double shock system and a CD}, \mpo{and the development time of the shocks is about $2\Omega_i^{-1}$. \jnn{In general}, the structure of the shocks is similar in all runs; they consist of the upstream, foot, overshoot, and downstream regions. In detail, the influence of the global magnetic field on the phase-space distribution of shock-reflected ions imposes differences in the evolution of the Buneman and the Weibel instability. Magnetic reconnection events are observed at the Weibel-mode zone} in simulations with $0^o$ and $45^o$ magnetic-field orientations.

\item The simulated shocks undergo cyclic self-reformation on account of the nonstationary character of ion reflection at the shock. The period of shock reformation is $\sim 1.55\Omega^{-1}$ which is close to the value obtained in \jnn{our} earlier investigations ($1.5\Omega^{-1}$ in \citet{2016ApJ...820...62W}). This period \mpon{neither depends on the magnetic-field configuration nor on} the plasma beta. 

\item Reflected ions interact with incoming electrons and excite electrostatic Buneman waves at the leading edge of the foot region. The parameters of our simulations satisfy the trapping condition \citep{2012ApJ...755..109M}, and so electrons can be  accelerated by the \jnn{convective electric} field. The intensity of the electrostatic waves, the area they occupy, and, subsequently, the fraction of nonthermal electrons strongly depend on the global magnetic-field orientation. Simulations with the out-of-plane magnetic-field configuration produce a much higher fraction of nonthermal electrons.
% than are found for $\varphi=45^o$ or $\varphi=0^o$. 
\mponn{In the cases $\varphi=0^o$ and $\varphi=45^o$ part of the counterstreaming of reflected ions proceeds in $z$ direction, but the $k_z$ component of the Buneman waves driven by it are not captured by the $x$-$y$ simulation grid. Consequently, the saturation level of the waves and their ability to trap electrons is less than for $\varphi=90^o$, for which we observe a higher intensity of Buneman waves at the shock foot and a larger volume coverage. With equation~\ref{trappingnew} we propose a revision to the trapping condition originally formulated by \citet{2012ApJ...755..109M}, that offers an explanation why some of the simulation presented here, and others described in the literature, have a low Buneman efficiency. 
If the new revised trapping condition (\ref{trappingnew}) is met, as in our simulation with $\varphi=90^o$, \jnr{then} the number of electrons undergoing SSA is large, at least as long as the plasma beta $\beta_\mathrm{p}\lesssim 1$. A similar modification can be applied to the driving condition, leading to equation~\ref{unstablenew}.}

\item Shock self-reformation leads to temporal variations in the electrostatic-wave intensity in the foot region, and consequently the electron energization in these regions becomes time-dependent. After crossing the Buneman-wave zone the fraction of nonthermal electrons does not change significantly. The cyclic shock reformation then imposes quasi-periodic fluctuations \mponn{in the temperature and density of bulk electrons as well as the density of high-energy electrons, that lead to variations in the nonthermal fraction of electrons. The amplitude of these variations is largest for out-of-plane magnetic field, where we also observe spatial correlations between the temperature and the density of bulk electrons that render homogenization of downstream electrons a slow process. That is not the case for runs with $\varphi=0^o$ and $\varphi=45^o$, and consequently we observe the fraction of nonthermal electrons in the downstream region decreasing with time.}
%Distance between maxima in a profiles well correlate with one calculated via shock velocity and period of self-reformation.

\item \mpo{For all simulated shocks we find suprathermal tails in the electron spectra in the downstream region, but the acceleration efficiency depends on the magnetic-field orientation. For all configurations the main acceleration process is through interaction with electrostatic waves in the Buneman wave zone at the shock foot, followed by further energization by turbulent magnetic structures and \jnn{in} the shock overshoot. The weight of the individual contributions by all these processes depends on the magnetic-field configuration. The cases $\varphi=0^o$ and $\varphi=45^o$ lead to the same acceleration processes, and we do not see any significant difference between them. In contrast, the $\varphi=90^o$ configuration provides a fundamentally different behavior. In this case the intensity of Buneman waves at the shock foot is higher, \abr{because the new trapping condition (Equation~\ref{trappingnew}) is satisfied}, and the waves are coherent and found in a larger region \abr{as the result of coherent ion reflection by magnetic field at the overshoot}. One consequence is that electrons gyrating in the foot region can cross the Buneman wave zone more than once, experience more shock-surfing acceleration, and reach a higher energy than they would with a $\varphi=0^o$ or $\varphi=45^o$ configuration, for which we do not observe such multiple interactions. 
The number of high-energy electrons (at $\gamma>1.5$) in the Buneman zone is correspondingly larger for $\varphi=90^o$ than it is for the other cases. In all simulations the Buneman instability serves as injector, i.e., electrons are accelerated to suprathermal energies by electrostatic waves at the leading edge of the foot region. The second stage of electron energization is second-order Fermi acceleration by interaction with magnetic filaments in simulations with $0^o$ or $45^o$ configuration, and it is adiabatic acceleration in the case of the out-of-plane magnetic field. 
For all simulations, the Lorentz factor of the most energetic electron is $\gamma \approx 9$, but the fraction of nonthermal electrons is more then 10 times larger for $\varphi=90^o$ than for the other configuration on account of the higher SSA efficiency in the foot region.}
\mpo{There is no clear trend between the number of high-energy electrons and the $\beta_{\rm p}$ value of the plasma into which the shock propagates. For $\varphi=90^o$ we observe more electrons above $\gamma=3$ in the \jnn{moderate}-$\beta_{\rm p}$ case than for the low $\beta_{\rm p}$. For the other simulations, that have at least part of the large-scale magnetic field in the simulation plane, we find the opposite trend. This issue will be the subject of a separate publication.}
\ab{However for a given magnetic-field orientation, we observe a higher nonthermal fraction at shocks propagating into \jnn{moderate}-$\beta_{\rm p}$ plasma than for low $\beta_{\rm p}$.}
%\ab{There is a clear correlation between SSA efficiency and number of energetic particles in downstream spectra Nonthermal fraction is larger in high beta cases but there is no clear dependence between SSA efficiency and }
%Summarize we can conclude that acceleration processes in runs A and C are different with some similarities which arise from global shock structure. For both configurations first stage of acceleration is the shock surfing acceleration in Buneman instability region, there are differences in intensity of electrostatic field strength, area of unstable region and, subsequently, fraction of nonthermal electrons. Then electrons undergo adiabatic acceleration (including betatron acceleration) and interact with overshoot region in a case of run C while in run A electrons undergo diffusive shock acceleration through interaction with magnetic structures and compressive regions.
\end{itemize}

   \begin{table*}
      \caption[]{\abr{Comparison of the electron pre-acceleration efficiency in the present and in other published simulations. }}
         \label{table-all-sim}
\centering
\begin{tabular}{lcccccccc}
\hline
\hline
\noalign{\smallskip}
Run   & $\varphi$ & $\beta_\mathrm{p}$  & Eq.~\ref{trapping} & Eq.~\ref{trappingnew} & Nonthermal population & Notes \\
\noalign{\smallskip}
\hline
\noalign{\smallskip}
A1; A2  & $0^o$   & 0.0005;0.5   &  Yes  &  No & Weak & \\
\cite{2010ApJ...721..828K}  & $0^o$   & 26   &  Yes  &  Yes &  Absent  &  High temperature of reflected ions\\
\cite{2015Sci...347..974M}  & $0^o$   & 0.5   &  Yes  &  No & Weak or absent &   \\
\noalign{\smallskip}
\hline
\noalign{\smallskip}
B1; B2  & $45^o$  & 0.0005;0.5   &  Yes  &  No  & Weak & \\
\cite{2016ApJ...820...62W}  & $45^o$  & 0.0015;0.015   & Yes  & Yes  & Weak or absent & Spectra far downstream   \\
\noalign{\smallskip}
\hline
\noalign{\smallskip}
C1; C2  & $90^o$  & 0.0005;0.5   &  Yes & Yes  & Strong & \\
\cite{2012ApJ...755..109M}(Run A)  & $90^o$  & 0.5   &  Yes  & Yes  & Strong &   \\
\cite{2012ApJ...755..109M}(Run B)  & $90^o$  & 0.5   &  No &  No   & Weak &   \\
\cite{2012ApJ...755..109M}(Run C)  & $90^o$  & 0.5  &  Yes &  Yes  & Strong &   \\
\cite{2012ApJ...755..109M}(Run D)  & $90^o$  & 4.5   &  Yes &  Yes  & Weak & Weak driving   \\
%\cite{2013PhRvL.111u5003M}  & $90^o$  & 0.5  &  Yes  &  Yes & Strong    \\
\noalign{\smallskip}
\hline
\end{tabular}
\smallskip
\tablecomments{\abr{We state whether or not the old (Eq.~\ref{trapping}) and the revised (Eq.~\ref{trappingnew}) trapping conditions are met. Notes are presented for simulation with weak nonthermal population and where both old (Eq.~\ref{trapping}) and the revised (Eq.~\ref{trappingnew}) trapping conditions are satisfied.}} 
   \end{table*}

This paper presents evidence for significant variation in the efficiency of electron acceleration at perpendicular high-Mach-number shocks, depending on the choice of orientation of the large-scale magnetic field with respect to the simulation plane. 
\mponn{Much but not all of the variation can be traced to the efficiency of driving the Buneman instability at the shock foot. Our findings are summarized in Table~\ref{table-all-sim} in the context of other published results \citep{2010ApJ...721..828K,2012ApJ...755..109M,2015Sci...347..974M,2016ApJ...820...62W} and in present article are summarized in Table~\ref{table-all-sim}. There are three parameters that have an influence on a downstream spectra and final fraction of nonthermal electrons: the plasma beta, $\beta_\mathrm{p}$, the trapping condition in earlier (\ref{trapping}) and in revised form (\ref{trappingnew}), and the driving condition that can be likewise revised (\ref{unstablenew}). For $\varphi=90^o$ case the old and new trapping conditions are identical, and the nonthermal electron population appears to be sparse if the trapping condition is not satisfied (run B in \citet{2012ApJ...755..109M}) or if a high plasma beta leads to early saturation the Buneman instability (The thermal velocity of electrons is about half the streaming speed of reflected ions in run D of \citet{2012ApJ...755..109M}). A weak population of nonthermal electrons is observed in our runs with $\varphi=0^o$ and $\varphi=45^o$ as well as in the simulation described by \citet{2015Sci...347..974M}, because of the Alfv\'enic Mach number is too low to satisfy the revised trapping condition. \citet{2016ApJ...820...62W} discuss shocks with Mach numbers $M_A$ and $M_s$ large enough to drive Buneman waves and trap electrons, even if the modified conditions are applied. Electron spectra are extracted in the far-downstream region where leakage of nonthermal electrons is observed (see section \ref{reformation-nonthermal}), which may explain why very few energetic electrons were observed. The modified driving and trapping conditions were also met in the simulation of \citet{2010ApJ...721..828K} \mponn{where $M_A$ is large enough to satisfy the modified trapping condition}, but the authors argue that the high temperature of reflected ions reduced the Buneman growth rate by about an order of magnitude. Further studies of the non-uniformity of ion reflection from a corrugated overshoot structure are needed to confirm this explanation.}

We have presented results of 2D3V simulations, but the real world is 3D throughout. The question arises which behaviour one would observe in 3D and which of the 2D3V configurations provides the closest match to the 3D case. By definition, all wavevectors lie in the simulation plane, and so for $\varphi=90^o$ only electromagnetic modes with $B_x$ or $B_y$ components can rotate particles out of the simulation plane. Particle trajectories are thus approximately confined to the simulation plane, and the particle ensemble assumes an effective adiabatic index of $2$, as opposed to $5/3$ for a 3D monoatomic gas. We indeed observe the corresponding difference in shock speeds, etc., \mponn{and the 3D shock structure may be not accurately reproduced for $\varphi=90^o$. On the other hand, with $\varphi=90^o$ the counterstreaming of reflected ions and the driving of Buneman modes in the shock foot is fully captured by the $x$-$y$ simulation grid. The fair fraction of nonthermal electrons produced in $\varphi=90^o$ simulations is probably a better indicator of the 3D acceleration efficiency at the shock foot than is the very low abundance of energetic electrons for $\varphi=0^o$ and $\varphi=45^o$.} True 3D simulations are urgently needed to resolve this issue.

---------------------------

\acknowledgments
%The authors acknowledge the Texas Advanced Computing Center (TACC) at The University of Texas at Austin for providing HPC and visualization resources that have contributed to the research results reported within this paper. We are also grateful for HPC resources provided by The North-German Supercomputing Alliance (HLRN). 
The work of A.B., J.N. and O.K. is supported by Narodowe Centrum Nauki through research project DEC-2013/10/E/ST9/00662.
M.P. acknowledges support through grants PO 1508/1-1 and PO 1508/1-2 of the Deutsche Forschungsgemeinschaft. 
\jnn{Numerical simulations have been performed on the Prometheus system at ACC
Cyfronet AGH.} \mpo{Part of the numerical work was \jnn{also} conducted on resources provided by The North-German Supercomputing Alliance (HLRN) under project bbp00003.}

\end{document}